\documentclass[preprint]{JHEP3} % 10pt is ignored!
\usepackage{epsfig,multicol}
\usepackage{times}

\usepackage{epsf}
\def\beq{\begin{equation}}
\def\eeq{\end{equation}}
\def\bea{\begin{eqnarray}}
\def\eea{\end{eqnarray}}

\def\nnb{\nonumber}

\def\rar{\rightarrow}
\def\nnb{\nonumber}

\def\ba{\begin{array}}
\def\ea{\end{array}}
\def\bea{\begin{eqnarray}}
\def\eea{\end{eqnarray}}
\def\Brll{$B\rightarrow \rho \ell^+ \ell^-$}
\def\Bpll{$B\rightarrow \pi \ell^+ \ell^-$}
\def\Bptt{$B\rightarrow \pi \tau^+ \tau^-$}
\def\Brtt{$B\rightarrow \rho \tau^+ \tau^-$}
\def\tep{$b \rar d \ell^+ \ell^-$}

\newcommand\fverb{\setbox\pippobox=\hbox\bgroup\verb}
\newcommand\fverbdo{\egroup\medskip\noindent%
            \fbox{\unhbox\pippobox}\ }
\newcommand\fverbit{\egroup\item[\fbox{\unhbox\pippobox}]}
\newbox\pippobox
\title{The exclusive \Bpll and \Brll decays in the general two Higgs doublet model}

\author{G\"{u}ray Erkol~
        and G\"{u}rsevil Turan\\
    Middle East Technical University, Inonu Bul. $06531$ Ankara, Turkey \\
    E-mail: \email{gurerk@newton.physics.metu.edu.tr}, \email{gsevgur@metu.edu.tr}}
\abstract{We study the differential branching ratio, branching ratio and the
forward-backward asymmetry for the exclusive \Bpll and \Brll decays in
the general two Higgs doublet model including the neutral Higgs
boson effects. We analyze the dependencies  of these quantities on the neutral Higgs boson contributions and the other model parameters. We observe that 
two Higgs doublet model with the neutral Higgs boson exchanges gives quite sizable contributions to these observables for both channels we consider. Since the neutral Higgs boson exchanges are the only source of the
forward-backward asymmetry for \Bptt decay, which is at the order of magnitude $1-10\%$,  measurement of  this observable is promising to determine the neutral Higgs boson effects.}

\begin{document}

\section{Introduction}
The rare decays of B-mesons, induced by the flavor-changing neutral
currents (FCNC), have always been a good candidate for testing the
Standard Model (SM) at the loop level and looking for new physics
beyond it. They can also be used to determine the fundamental
parameters of the SM, like the elements of the
Cabibbo-Kobayashi-Maskawa (CKM) matrix,  the leptonic decay
constants etc.. Among the rare B-decays, exclusive processes
induced by $b\rightarrow s(d)\ell^+\ell^-$ transitions have
received a special attention since the SM predicts relatively
larger branching ratios for these decays.

From the experimental side, there is an impressive effort for
searching B-meson decays, especially  at the B-factories, like at Belle
\cite{Belle} and BaBar \cite{Babar}, and with the increased statistical
power of these experiments, in the near future the
rare B-meson decays will be measured very precisely. From the
theoretical side, the decays $B\rightarrow X_{s,d} \ell^+ \ell^-$
provide important probes of the effective Hamiltonian which
governs the FCNC transitions $b\rightarrow s(d)\ell^+\ell^-$ at
quark level \cite{AAli}. For $b\rightarrow s\ell^+\ell^-$
transition, the matrix element contains the terms that receive
contributions from $t\bar{t}$, $c\bar{c}$ and $u\bar{u}$ loops,
which are proportional to the combination of
$\xi_t=V_{tb}V^*_{ts}$, $\xi_c=V_{cb}V^*_{cs}$ and
$\xi_u=V_{ub}V^*_{us}$, respectively. Smallness of $\xi_u$
in comparison with $\xi_c$ and $\xi_t$, together with the
unitarity of the CKM matrix elements, bring about the consequence
that matrix element for the $b\rightarrow s\ell^+\ell^-$ decay
involves only one independent CKM factor $\xi_t$, so that the
CP violation in this channel is suppressed  in the SM \cite{Aliev00,Du}.
However, for $b\rightarrow d\ell^+\ell^-$ decay, all the CKM
factors $\eta_t=V_{tb}V^*_{td}$, $\eta_c=V_{cb}V^*_{cd}$ and
$\eta_u=V_{ub}V^*_{ud}$ are at the same order in the SM and
this induces a considerable CP violating asymmetry in the partial
rates \cite{Kruger1, Kruger2}.

In this paper we investigate the exclusive \Bpll and \Brll decays,
which are induced by the \tep decay at the quark level, in the
framework of the general two Higgs doublet model (2HDM) (model III). These decays have been
studied in the literature, both in the SM and in
the 2HDM. CP violating effects in inclusive $b\rightarrow
d\ell^+\ell^-$ and exclusive \Bpll and \Brll channels were studied
within the framework of the SM in refs.
\cite{Kruger1}-\cite{Bertoluni}. 2HDM contributions to  these
exclusive decays have been investigated in \cite{Aliev1,Iltan1}.
In earlier works about the exclusive \Brll and \Bpll decays,
contributions from exchanging the neutral
Higgs bosons  (NHB) were neglected because of the
smallness of $m_{\ell}/m_W \, (\ell=e,\mu)$. However, in the
models with two Higgs doublets, such as MSSM, 2HDM, etc., the
situation is different, especially in case of $\ell=\tau$ with
large $\tan \beta$ or large $\bar{\xi}^U_{N,\tau \tau}$, where
$\tan \beta=v_2/v_1$, the ratio of the vacuum expectation values of
the two Higgs doublets and $\bar{\xi}^U_{N,\tau \tau}$ is the
Yukawa coupling to which NHB contributions are proportional  in
model III. Indeed, there are a number of works in the literature, which
show that the contributions from exchanging NHB can compete with
those from exchanging $\gamma$ and Z when $\tan \beta$ and/or
$\bar{\xi}^U_{N,\tau \tau}$ are large enough \cite{Iltan2}-\cite{Xiong}.

The aim of this work is to calculate the branching ratio ($BR$)
and the forward backward asymmetry ($A_{FB}$) of the exclusive
\Bpll and \Brll decays in the general 2HDM, including NHB effects without 
neglecting the lepton mass. The
2HDM is one of the simplest extensions of the SM, which is
obtained  by the addition of a second complex Higgs doublet. In
general, the 2HDM possesses tree-level FCNC that can be avoided by
imposing an {\it ad hoc} discrete symmetry \cite{Glashow}. As a
result, there appear two different choices, namely model I and II,
depending on whether up-type and down-type quarks couple to the
same or two different Higgs doublets, respectively. Model II has
been more attractive since its Higgs sector is the same as the
Higgs sector in the supersymmetric models. In a more general 2HDM,
namely model III \cite{modelIII,Soni}, no discrete symmetry is
imposed and there appear FCNC naturally at the tree level. We note that
in model III, FCNC receiving contributions from the first two
generations are highly suppressed, which is confirmed by the low
energy experiments. As for those involving the third generation,
it is possible to impose some restrictions on them with the
existing experimental results.

The paper is organized as follows: In section \ref{s2}, after we present
the theoretical framework of the general 2HDM and the leading order QCD corrected effective Hamiltonian for the process \tep, we calculate the differential $BR$ and the $A_{FB}$ of the exclusive \Brll and \Bpll decays. 
The \ref{s3}. section is devoted to the numerical analysis and the discussions.
Finally, in the Appendices, we give the explicit forms of the operators appearing in the effective Hamiltonian and the corresponding Wilson coefficients.

\section{The exclusive $B \rar \pi \ell^+ \ell^- $ and \Brll decays in the
framework of the general 2HDM \label{s2}}

 In this section, we first present
the theoretical framework of the general 2HDM, and then calculate
some physical observables  related to the exclusive \Bpll and
\Brll decays in this model.
\subsection{The theoretical framework}

We would like to present the main essential points of the general
2HDM, namely model III. In this model, both Higgs doublets can
couple to up- and down-types quarks. We can choose two scalar
doublets $\phi_1$ and $\phi_2$ in the following form
\begin{eqnarray}
\phi_{1}=\frac{1}{\sqrt{2}}\left[\left(\begin{array}{c c}
0\\v+H^{0}\end{array}\right)\; + \left(\begin{array}{c c} \sqrt{2}
\chi^{+}\\ i \chi^{0}\end{array}\right) \right]\, ;
\phi_{2}=\frac{1}{\sqrt{2}}\left(\begin{array}{c c} \sqrt{2}
H^{+}\\ H_1+i H_2 \end{array}\right) \,\, \label{choice}
\end{eqnarray}
with the vacuum expectation values,
\begin{eqnarray}
<\phi_{1}>=\frac{1}{\sqrt{2}}\left(\begin{array}{c c}
0\\v\end{array}\right) \,  \, ; <\phi_{2}>=0 \,\,
\label{choice2}
\end{eqnarray}
so that the first doublet $\phi_1$ is the same as the one in the
SM, while the second doublet contains all the new particles.
Further, we take $H_1$ and $H_2$ as the mass eigenstates $h^0$ and
$A^0$, respectively.

The general Yukawa Lagrangian can be written as
\begin{eqnarray}
{\cal{L}}_{Y}&=&\eta^{U}_{ij} \bar{Q}_{i L} \tilde{\phi_{1}} U_{j
R}+ \eta^{D}_{ij} \bar{Q}_{i L} \phi_{1} D_{j R}+ \xi^{U\,
\dagger}_{ij} \bar{Q}_{i L} \tilde{\phi_{2}} U_{j R}+ \xi^{D}_{ij}
\bar{Q}_{i L} \phi_{2} D_{j R} + h.c. \,\,\, , \label{lagrangian}
\end{eqnarray}
where $i,j$  are family indices of quarks , $L$ and $R$ denote
chiral projections $L(R)=1/2(1\mp \gamma_5)$, $\phi_{m}$ for
$m=1,2$, are the two scalar doublets, $Q_{i L}$ are quark
doublets, $U_{j R}$, $D_{j R}$ are the corresponding quark
singlets, $\eta^{U,D}_{ij}$ and $\xi^{U,D}_{ij}$ are the matrices
of the Yukawa couplings. After the rotation that diagonalizes the quark 
mass eigenstates, the part of the Lagrangian that is responsible for the FCNC at the
tree level looks like
\begin{eqnarray}
{\cal{L}}_{Y,FC}= -H^{\dagger} \bar{{\cal{U}}}[V_{CKM}\, \xi^{D}_{N}\, 
R-\xi^{U,\dagger}_{N}\, V_{CKM} \, L] \,  {\cal{D}}   \,\, ,
\label{lagrangianFC}
\end{eqnarray}
where ${\cal{U}}({\cal{D}})$ represents the mass eigenstates of up (down) type quarks.
In this work, we adopt the following
redefinition of the Yukawa couplings:
\bea \xi^{U,D}_N=\sqrt{\frac{4G_F}{\sqrt{2}}}\bar{\xi}^{U,D}_{N,ij}. \eea

After this brief summary about the general 2HDM, now we would like
to present briefly the main steps in calculating the matrix
elements for the inclusive $b\rightarrow d \ell^+ \ell^-$ decay.
For this, the effective Hamiltonian method provides a powerful
framework. In this approach, the first step is to calculate the
full theory including the NHB effects. We use the on-shell
renormalization scheme to overcome the logarithmic divergences
that appear during the calculations of NHB contributions. (For the
details of this calculations, see ref.\cite{Iltan2}.) The next
step is to match the full theory with the effective theory, which
is obtained by integrating out the heavy degrees of freedom, i.e.,
$t$ quark, $W^{\pm}$, $H^{\pm}$, $h^0$, $H^0$ and $A^0$ in our
case, at high scale $\mu=m_W$, and then calculate the Wilson
coefficients at the lower scale $\mu \sim {\cal O}( m_b)$ using
the renormalization group equations. Following these steps above,
one can obtain the effective Hamiltonian governing the $b
\rightarrow d \ell^+ \ell^-$ transitions, in the 2HDM in terms of
a set of operators 
\begin{eqnarray}\label{Hamiltonian}
{\cal H}_{eff} & = & \frac{4 G_F}{\sqrt{2}} \, V_{tb}
V^*_{td}\Bigg\{ \sum_{i=1}^{10} \, \, C_i (\mu ) \, O_i(\mu) \,
+\sum_{i=1}^{10} \, \, C_{Q_i} (\mu ) \, Q_i(\mu) \, \nnb\\
&-&\lambda_u
\{C_1(\mu)[O_1^u(\mu)-O_1(\mu)]+C_2(\mu)[O_2^u(\mu)-O_2(\mu)]\}\Bigg\}
\end{eqnarray}
where \bea\label{CKM}
\lambda_u=\frac{V_{ub}V_{ud}^\ast}{V_{tb}V_{td}^\ast}, \eea using
the unitarity of the CKM matrix i.e.
$V_{tb}V_{td}^\ast+V_{ub}V_{ud}^\ast=-V_{cb}V_{cd}^\ast$. Here,
$O_1$ and $O_2$ are \emph{the current-current operators},
$O_3$,...,$O_6$ are usually named as \emph{the QCD penguin
operators}, $O_7$ and $O_8$ are \emph{the magnetic penguin
operators} and $O_9$ and $O_{10}$ are \emph{the semileptonic
electroweak penguin operators}. The additional $Q_i$
($1=1,...,10$) are due to the NHB exchange diagrams.
$O_1^u$-$O_2^u$ are the new operators for the $b\rar d$ decay
which are absent in the $b\rar s$ decay. $C_i(\mu)$ and
$C_{Q_i}(\mu)$ are Wilson coefficients renormalized at the scale
$\mu$. All these operators and the Wilson coefficients, together
with their initial values calculated at $\mu=m_{W}$ in the SM and
also additional coefficients coming from the new Higgs scalars are
presented in Appendix \ref{App1}.

Neglecting the mass of the $d$ quark, the effective short distance
Hamiltonian for the $b \rightarrow d \ell^+ \ell^-$ decay leads
to the QCD corrected matrix element:
\begin{eqnarray}\label{genmatrix}
{\cal M} &=&\frac{G_{F}\alpha}{2\sqrt{2}\pi }V_{tb}V_{td}^{\ast }%
\Bigg\{C_{9}^{eff}(m_{b})~\bar{d}\gamma _{\mu }(1-\gamma _{5})b\,\bar{\ell}%
\gamma ^{\mu }\ell +C_{10}(m_{b})~\bar{d}\gamma _{\mu }(1-\gamma _{5})b\,\bar{%
\ell}\gamma ^{\mu }\gamma _{5}\ell  \nonumber \\
&-&2C_{7}^{eff}(m_{b})~\frac{m_{b}}{q^{2}}\bar{d}i\sigma _{\mu \nu
}q^{\nu }(1+\gamma _{5})b\,\bar{\ell}\gamma ^{\mu }\ell
+C_{Q_1}(m_b)~\bar{d}(1+\gamma_5)b\,\bar{\ell}\ell+C_{Q_2}(m_b)~\bar{d}(1+\gamma_5)b\,\bar{\ell}
\gamma_5\ell\Bigg\},\nonumber\\
\end{eqnarray}
where $q$ is the momentum transfer.

\subsection{The exclusive \Bpll decay in the 2HDM}
In this section we calculate the $BR$ and the $A_{FB}$ of the
\Bpll decay. In order to find these physically measurable
quantities at hadronic level, we first need to calculate the
matrix elements
$<\pi(p_\pi)|\bar{d}\gamma_\mu(1-\gamma_5)b|B(p_B)>$,
$<\pi(p_\pi)|\bar{d}i\sigma_{\mu\nu}q_\nu(1+\gamma_5)b|B(p_B)>$
and $<\pi(p_\pi)|\bar{d}(1+\gamma_5)b|B(p_B)>$. The first two of
these matrix elements can be written in terms of the form factors
in the following way

\bea\label{matrixp1}
<\pi(p_\pi)|\bar{d}\gamma_\mu(1-\gamma_5)b|B(p_B)>&=&f^+(q^2)(p_B+p_\pi)_\mu
+f^-(q^2)q_\mu\, ,\eea \bea\label{matrixp2}
<\pi(p_\pi)|\bar{d}i\sigma_{\mu\nu}q_\nu(1+\gamma_5)b|B(p_B)>&=&[(p_B+p_\pi)_\mu
q^2-q_\mu(m_B^2-m_\pi^2)]f_v(q^2)\, ,
\eea
where $p_B$ and $p_\pi$ denote the four momentum vectors of $B$
and $\pi$-mesons, respectively. To find
$<\pi(p_\pi)|\bar{d}(1+\gamma_5)b|B(p_B)>$, we multiply both sides
of  Eq. (\ref{matrixp1}) with  $q_{\mu}$ and then use the equation
of motion. Neglecting the mass of the $d$-quark, we get
\bea\label{matrixp3}
<\pi(p_\pi)|\bar{d}(1+\gamma_5)b|B(p_B)>=\frac{1}{m_b}[f^+(q^2)(m_B^2-m_\pi^2)+f^-(q^2)
q^2 ].
\eea
Using  Eqs. (\ref{matrixp1}-\ref{matrixp3}), we  find the
amplitude governing the \Bpll decay :
\bea\label{matrixp}
\cal{M}^{B\rightarrow\pi}&=&\frac{G_F\alpha}{2\sqrt{2}\pi}V_{tb}V_{td}^\ast\Bigg
\{[2A_1 p_\pi^\mu+B_1 q^\mu]\bar{\ell}\gamma_\mu \ell+[2G_1
p_\pi^\mu +D_1 q^\mu]\bar{\ell}\gamma_\mu\gamma_5\ell
+E_1\bar{\ell}\ell+F_1 \bar{\ell}\gamma_5 \ell \Bigg\}\nnb\\
\eea
where
\bea A_1&=&C^{eff}_9 f^+ -2m_B C^{eff}_7 f_v,\nnb\\
B_1&=&C^{eff}_9(f^+ +f^-)+2 C^{eff}_7
\frac{m_B}{q^2}f_v(m_B^2-m_\pi^2-q^2),\nnb\\
G_1&=&C_{10}f^+,\nnb\\
D_1&=&C_{10}(f^+ +f^-),\nnb\\
E_1&=&C_{Q_1}\frac{1}{m_b}[(m_B^2-m_\pi^2)f^+ +f^- q^2],\nnb\\
F_1&=&C_{Q_2}\frac{1}{m_b}[(m_B^2-m_\pi^2)f^+ +f^- q^2]. \eea Here
$f^+$, $f^-$ and $f_v$ are the relevant form factors.

The matrix element in Eq. (\ref{matrixp}) leads to the following
double differential decay rate:
\begin{eqnarray}\label{ddrp}
\frac{d^2\Gamma}{ds ~dz}&=&\frac{G_F^2
\alpha^2}{2^{11}\pi^5}|V_{tb}V_{td}^\ast|^2 m_B^3 \sqrt{\lambda_\pi}~v~
\Bigg\{m_B^2 \lambda_{\pi}(1-z^2 v^2)|A_1|^2+ s(v^2 |E_1|^2+ |F_1|^2) \nonumber
\\&+& (m_B^2 \lambda_{\pi}(1-z^2 v^2) +16~r_\pi~m^2_{\ell})~|G_1|^2
+4~s~m_{\ell}^2~|D_1|^2\nonumber\\
&+& 4~m_{\ell}^2~(1-r_{\pi}-s)~Re[G_1D_1^\ast]+2~v~m_{\ell}~\sqrt{\lambda_{\pi}}~z~
Re[A_1E_1^\ast]\nonumber\\
&+&2~m_{\ell}~((1-r_{\pi}-s)~Re[G_1F_1^\ast]+s Re[D_1F_1^\ast])
\Bigg\} \, .\end{eqnarray}
Here $s=q^2/m_B^2$, $r_\pi=m_\pi^2/m_B^2$, $v=\sqrt{1-\frac{4t^2}{s}}$, $t=m_l/m_B$, 
$\lambda_{\pi}=r_\pi^2+(s-1)^2-2r_\pi(s+1)$ and $z=\cos\theta$, 
where $\theta$ is the angle between the three-momentum of the $\ell^-$ lepton and that 
of the B-meson in the center of mass frame of the dileptons $\ell^+\ell^-$. We note that 
our expression for double differential decay rate in Eq. (\ref{ddrp}) coincides with the 
one in \cite{Erhanx}.

Integrating the expression in Eq. (\ref{ddrp}) over the angle variable, we obtain  the
 differential decay rate as follows 
\bea
\frac{d\Gamma}{ds}&=&\frac{G_F^2
\alpha^2}{2^{10}\pi^5}|V_{tb}V_{td}^\ast|^2 m_B^3 \sqrt{\lambda_\pi} \, v \, \Delta_{\pi} \, ,
\eea
where
\bea
\Delta_{\pi} & = & \frac{1}{3}~m_B^2~\lambda_{\pi}(3-v^2)(|A_1|^2+|G_1|^2)
+\frac{4m_{\ell}^2}{3s}(12\,r_\pi~s+\lambda_\pi)|G_1|^2\nnb\\&+&4~m_{\ell}^2~s~|D_1|^2+
s (v^2|E_1|^2+|F_1|^2)+4~m_{\ell}^2 (1-r_{\pi}-s) Re[G_1~D_1^\ast]\nnb\\
&+& 2~m_{\ell}((1-r_{\pi}-s)Re[G_1~F_1^\ast]+s Re[D_1~F_1^\ast]) \, .\label{deltapi}
\eea

The $A_{FB}$ is another observable that may provide valuable
information at hadronic level. We write its definition as given by
\begin{eqnarray}\label{genafb}
A_{FB}(s)& = & \frac{ \int^{1}_{0}dz \frac{d \Gamma }{dz} -
\int^{0}_{-1}dz \frac{d \Gamma }{dz}}{\int^{1}_{0}dz \frac{d
\Gamma }{dz}+ \int^{0}_{-1}dz \frac{d \Gamma }{dz}} \, ,
\end{eqnarray}
where $\Gamma$ is the total decay rate.

For the $B \rar \pi \ell^+ \ell^- $ decay,  $A_{FB}$  is
calculated to be
\bea A_{FB}&=&-\int\,\,ds\,\, (t v^2 \lambda_{\pi}
Re(A_1~E_1^\ast))\Bigg/ \int\,\,ds\,\, v \sqrt{\lambda_{\pi}} \, \Delta_{\pi}
.\label{piafb} \eea
As seen from Eq.(\ref{piafb}), in the $B \rar \pi \ell^+ \ell^- $
decay, the only source for the $A_{FB}$ is the NHB effects  \cite{Kruger3,Erhanx}.
Since $A_{FB}$ does not exist in the SM and 2HDM without NHB effects, it
may be a good candidate for testing the existence and the
importance of the NHB contributions .

\subsection{The exclusive \Brll decay in the 2HDM}
In this section we proceed to calculate the $BR$ and the $A_{FB}$
of the \Brll decay. We follow the same strategy as in the \Bpll
decay. In order to calculate the matrix element governing the
\Brll decay, we need the following matrix elements:
\bea\label{matrix1}
<\rho(p_\rho,\varepsilon)|\bar{d}\gamma_{\mu}(1-\gamma_5)b|B(p_B)>&=&
-\epsilon_{\mu\nu\lambda\sigma}\varepsilon^{\ast\nu}p_
{\rho}^\lambda
p_B^\sigma\frac{2V(p^2)}{m_B+m_\rho}-i\varepsilon_{\mu}^\ast
(m_B+m_\rho)A_1(q^2)\nnb\\
&+&i(p_B+p_\rho)_\mu (\varepsilon^\ast
q)\frac{A_2(q^2)}{m_B+m_\rho}+i q_\mu(\varepsilon
q)\frac{2m_\rho}{q^2}[A_3(q^2)\nnb\\&&-A_0(q^2)],\eea

\bea\label{matrix2}
<\rho(p_\rho,\varepsilon)|\bar{d}i\sigma_{\mu\nu}q^\nu(1+\gamma_5)b|B(p_B)>&=&
4\epsilon_{\mu\nu\lambda\sigma}
\varepsilon^{\ast\nu}p_{\rho}^\lambda
q^{\sigma}T_1(q^2)+2i[\varepsilon^\ast_\mu(m_B^2-m_\rho^2)\nnb\\
&-&(p_B+p_\rho)_\mu(\varepsilon^\ast
q)]T_2(q^2)+2i(\varepsilon^\ast q)\nnb\\&&
\Bigg(q_\mu-(p_B+p_\rho)_\mu\frac{q^2}{m_B^2-m_\rho^2}\Bigg)T_3(q^2),
\eea
\bea\label{matrix3}
<\rho(p_\rho,\varepsilon)|\bar{d}(1+\gamma_5)b|B(p_B)>=\frac{-1}{m_b}2im_{\rho}(\varepsilon
q)A_0\, , \eea
where $p_\rho$ and $\varepsilon^\ast$ denote the four momentum and
polarization vectors of the $\rho$ meson, respectively. In order to calculate
the matrix element in Eq. (\ref{matrix3}), we multiply both sides of
Eq.(\ref{matrix1}) with  $q_{\mu}$ and use the equation of motion.

From Eqs. (\ref{matrix1}-\ref{matrix3}), we get the following
expression for the matrix element of   the \Brll decay:
\bea
\label{matrixBrll}\cal{M}^{B\rightarrow\rho}&=&\frac{G_F
\alpha}{2\sqrt{2}\pi}V_{tb}V_{td}^\ast\Bigg \{
\bar{\ell}\gamma_\mu\ell[2A\epsilon_{\mu\nu\lambda\sigma}
\varepsilon^{\ast\nu} p_\rho^\lambda p_B^\sigma +i B
\varepsilon^\ast_{\mu}-i C(p_B+p_\rho)_\mu (\varepsilon^\ast q)-i
D
(\varepsilon^\ast q)q_\mu]\nnb\\
&+& \bar{\ell}\gamma_\mu \gamma_5 \ell[2E
\epsilon_{\mu\nu\lambda\sigma}\varepsilon^{\ast\nu} p_\rho^\lambda
 p_B^\sigma +i F \varepsilon^\ast_{\mu} -i G(\varepsilon^\ast q)(p_B+p_\rho)
-i H(\varepsilon^\ast q) q_\mu]+i \bar{\ell}\ell
Q(\varepsilon^\ast q)\nnb\\&+&i \bar{\ell}\gamma_5 \ell N
(\varepsilon^\ast q)\Bigg \} \eea where
\bea A&=&C^{eff}_9\frac{V}{m_B+m_\rho}+4\frac{m_b}{q^2}C^{eff}_7 T_1,\nnb\\
B&=&(m_B+m_\rho)\Bigg( C^{eff}_9 A_1+\frac{4 m_b}{q^2}(m_B-m_\rho)C^{eff}_7
T_2\Bigg),\nnb\\
C&=&C^{eff}_9\frac{A_2}{m_B+m_\rho}+
4\frac{m_b}{q^2}C^{eff}_7\Bigg(T_2+\frac{q^2}{m_B^2-m_\rho^2}T_3\Bigg),\nnb\\
D&=&2C^{eff}_9\frac{m_\rho}{q^2}(A_3-A_0)-4C^{eff}_7\frac{m_b}{q^2} T_3,\nnb\\
E&=&C_{10} \frac{V}{m_B+m_\rho},\\
F&=&C_{10}(m_B+m_\rho)A_1,\nnb\\
G&=&C_{10}\frac{A_2}{m_B+m_\rho},\nnb\\
H&=&2C_{10}\frac{m_\rho}{q^2}(A_3-A_0),\nnb\\
Q&=&2C_{Q_1}\frac{m_\rho}{m_b}A_0,\nnb\\
N&=&2C_{Q_2}\frac{m_\rho}{m_b}A_0.\nnb \eea Here $A_0$, $A_1$,
$A_2$, $A_3$, $V$, $T_1$, $T_2$ and $T_3$ are the relevant form factors.

This matrix element leads to the following double differential
decay rate
\bea\label{ddrate}
\frac{d^2\Gamma}{ds~dz}&=&\frac{\alpha^2 G_F^2}{2^{15} m_B
\pi^5}|V_{tb} V^*_{td}|^2 \sqrt{\lambda_\rho}~v~\Bigg \{
4~s~\lambda_\rho (2+v^2(z^2-1)) |A|^2
+4\, v^2 \,s\, m_B^4 \lambda_\rho(1+z^2) |E|^2\nonumber\\
&+&16\, m_B^2\, s \,v\,  z\sqrt{\lambda_\rho}\Big(Re[B E^\ast]+Re[A F^\ast]\Big)
 + \frac{1}{r}\Bigg[[\lambda_\rho (1-z^2 v^2)+2\, r_\rho s (5-2 v^2)]|B|^2\nonumber\\
&+&m_B^4\lambda_\rho^2 (1-z^2 v^2)|C|^2+
[\lambda_\rho (1-z^2 v^2)-2\, r_\rho s (1-4 v^2)]|F|^2 \nonumber\\
&+&m_B^4\lambda_\rho[(-1+r_\rho)^2(1-v^2) z^2+
(-1+z^2)(s t^2-8(1+r_\rho)t^2-\lambda_\rho)]|G|^2\nonumber\\
&+&2\,m_B^2 \lambda_\rho W_\rho(1-z^2 v^2) Re[B C^\ast]-
2\, m_B^2 \lambda_\rho [W_\rho (1-z^2 v^2)-4t^2]Re[FG^\ast]\nonumber\\
&+&m_B^2 \lambda_\rho\Big(4 s m_{\ell}(m_{\ell}|H|^2+Re[H N^\ast])+s(|N|^2+v^2 |Q|^2)
-4 t (Re[F(2 t H^\ast+N^\ast/m_B)]\nnb \\ &+&4 (1-r_{\rho})m_{\ell}
Re[G(2 m_{\ell}H^\ast+N^\ast)]\Big)
+ 4 t m_B v z^2 
Re[(W_{\rho} B+m^2_B (W^2_{\rho}-4 r_{\rho}s) C)Q^\ast]\Bigg]\Bigg\}\, ,
\nnb \\ &&
\eea
where $r_{\rho}=m_\rho^2/m_B^2$, $W_\rho=-1+r_\rho + s$ and
$\lambda_\rho=r_{\rho}^2+(s-1)^2-2r_{\rho}(s+1)$.

The differential decay rate for \Brll decay is given by \cite{Cakmak}
\bea
\frac{d\Gamma}{ds}&=&\frac{\alpha^2 G_F^2 m_B}{2^{12} 
\pi^5}|V_{tb} V^*_{td}|^2 \sqrt{\lambda_\rho}~~v~~\Delta_{\rho}
\eea
where
\bea
\Delta_{\rho} & = & \frac{8}{3}\lambda_\rho m_B^6 s ((3-v^2)|A|^2+ 2 v^2 |E|^2)
-\frac{4}{r}\lambda_\rho m_B^2 m_{\ell}Re[(F-m_B^2 (1-r_\rho)G-m_B^2 s H)N^\ast]\nnb \\
&+& \frac{1}{r_\rho}\lambda_\rho m_B^4 \Bigg [ s v^2 |Q|^2+\frac{1}{3}\lambda_\rho m_B^2
(3-v^2)|C|^2+s |N|^2 +m_B^2s^2 (1-v^2)|H|^2 \nnb \\ 
& + & \frac{2}{3}[(3-v^2)\, W_\rho-3 \,s (1-v^2)] Re[F~G^\ast]-2\, s \,(1-v^2)Re[F~H^\ast]
\nnb \\ &+&2 \,m_B^2 s (1-r_\rho)(1-v^2)Re[G~H^\ast]+
\frac{2}{3}(3-v^2)W_\rho Re[B~C^\ast] \Bigg ]\nnb \\
& + & \frac{1}{3 r_\rho} m_B^2 \Bigg [ (\lambda_\rho +12 r_\rho s)(3-v^2)|B|^2
+\lambda_\rho m_B^4 [\lambda_\rho (3-v^2)-3 s (s-2 r_\rho-2)(1-v^2)]|G|^2\nnb \\
& + & (\lambda_\rho (3-v^2)+24 r_\rho s v^2)|F|^2 \Bigg] \label{deltarho}.
\eea

We also give the $A_{FB}$ of the \Brll decay 
\bea\label{afbrho}
A_{FB}&=&\int~ds~2m_B^3\lambda_\rho v^2 \Big( 4m_Bs(Re[B~E^\ast]+Re[A~F^\ast])\nnb\\
&+&\frac{t}{r_{\rho}}[W_\rho Re[B~Q^\ast]+m_B^2\lambda_\rho Re[C~Q^\ast]]\Big)\Bigg/\int~ds
\sqrt{\lambda_\rho} ~~v~~ \Delta_{\rho}.
 \eea

\section{Numerical results and discussion \label{s3}}
In this section we present the numerical analysis of the exclusive
\Bpll and \Brll decays in the general 2HDM. We give our results
for the $\ell=\tau$ case in order to express our motivation to
calculate the $BR$ and the $A_{FB}$ of these decays without
neglecting the lepton mass. The input parameters we used in our
numerical analysis are as follows:
\begin{eqnarray}
& & m_B =5.28 \, GeV \, , \, m_b =4.8 \, GeV \, , \,m_c =1.4 \,
GeV \, , \,
m_{\tau} =1.78 \, GeV \, , \, m_{\pi}=0.14 \, GeV \, ,  \nnb \\
& & m_{\rho}=0.77 \, GeV \, , \, m_{H^0} =150 \, GeV \, , \,m_{h^0} =100 \, GeV
\, , \,m_{A^0} =100 \, GeV \, , \, m_{H^{\pm}} =400 \, GeV \, , \nnb \\
& & |V_{tb} V^*_{td}|=0.011 \, , \, \alpha^{-1}=129 \,  ,
\,G_F=1.17 \times 10^{-5}\, GeV^{-2} \,  , \,\tau_B=1.54 \times
10^{-12} \, s \,  .
\end{eqnarray}
Using the Wolfenstein parametrization of the CKM matrix,
$\lambda_u$ in Eq.(\ref{CKM}) can be written as: \bea
\lambda_u=\frac{\rho(1-\rho)-\eta^2-i\eta}{(1-\rho)^2+\eta^2}+O(\lambda^2).
\eea
Furthermore, we have used the relation
\bea
\frac{|V_{tb} V_{td}^\ast|^2}{|V_{cb}|^2} & = & 
\lambda^2 [(1-\rho)^2+\eta^2]+{\cal O}(\lambda^4)
\eea
where $\lambda=\sin \theta_C\simeq 0.221$ and 
we take the Wolfenstein parameters as $\rho=-0.07$ and $\eta=0.34$ throughout 
the calculations.

We note that the Wilson coefficient $C_9^{eff}$ receives  also  long
distance (LD) contributions due to $c\bar{c}$ intermediate states.
(See Appendix \ref{App2} for the details of the LD contributions).\TABULAR{|c c c c|}
 {\hline\hline
   %after \\: \hline or \cline{col1-col2} \cline{col3-col4} ...
                  & $f(0)$ & $T_f$ & $n_f$ \\ \hline
  $f_{+}^{B\rar\pi}$ & 0.29 & 6.71 & 2.35 \\
  $f_{-}^{B\rar\pi}$ & -0.26 &6.553 & 2.30 \\
  $f_{v}^{B\rar\pi}$ & -0.05 & 6.68 & 2.31 \\
  \hline\hline}
 {\label{tabpi}$B\rar\pi$ transition form factors in the light-cone constituent quark model.}
There are five possible resonances in the $c\bar{c}$ system that
can contribute to the decays under consideration and to calculate
their contributions, we need to divide the integration regions for
$s$ into two parts: the region $0.4546 \leq s \leq
((m_{\psi_2}-0.02)/m_B)^2$ is common for both decays and we have
$((m_{\psi_2}+0.02)/m_B)^2 \leq s \leq 0.9476$ and
$((m_{\psi_2}+0.02)/m_B)^2 \leq s \leq 0.7296$ as second
integration parts for the \Bptt and \Brtt decays,
respectively. Here, $m_{\psi_2}$ is the mass of the second
resonance.
 \TABULAR{|c c c c|}
 {\hline\hline
   %after \\: \hline or \cline{col1-col2} \cline{col3-col4} ...
                  & $F(0)$ & $a_F$ & $b_F$ \\ \hline
  $A_1^{B\rar\rho}$ & $0.26\pm 0.04$ & 0.29 & -0.415 \\
  $A_2^{B\rar\rho}$ & $0.22\pm 0.03$ & 0.93 & -0.092 \\
  $V^{B\rar\rho}$ & $0.34\pm 0.05$ & 1.37 & 0.315 \\
  $T_1^{B\rar\rho}$ & $0.15\pm 0.02$ & 1.41 & 0.361 \\
  $T_2^{B\rar\rho}$ & $0.15\pm 0.02$ & 0.28 & -0.500 \\
  $T_3^{B\rar\rho}$ & $0.10\pm 0.02$ & 1.06 & -0.076 \\
  \hline\hline}
 {\label{tabrho}$B\rar\rho$ transition form factors in a
three-parameter fit.}

The masses of the charged and the neutral Higgs bosons,
$m_{H^\pm}$, $m_{A^0}$, $m_{h^0}$, $m_{H^0}$ and the Yukawa
couplings ($\xi_{ij}^{U,D}$) remain as free parameters of the
model. For the mass of the charged Higgs, the lower limit 
$m_{H_{\pm}}\geq 200$ GeV and $m_{H_{\pm}}\geq 250$ GeV  have
been given in \cite{ALEP} and \cite{Ciuchini}, respectively. However,
it is also pointed out that  adding different theoretical errors leads to
$m_{H_{\pm}}> 370$ GeV and also these bounds are quite sensitive to the details
of the calculations. In our work, we choose $m_{H_{\pm}}=400$ GeV . For the 
masses of the neutral Higgs bosons, the lower limits are given as
$m_{H^{0}}\geq 115$ GeV, $m_{h^{0}}\geq 89.9$ GeV $m_{A^{0}}\geq 90.1$ GeV
in \cite{CERN} and the values we choose are given in Eq.(3.1).  

For Yukawa couplings , we use the restrictions coming from
CLEO data \cite{CLEO},
\bea BR(B\rightarrow X_s \, \gamma)& = &
(3.15\pm 0.35\pm 0.32)10^{-4} \, ,\label{data}
\eea
$B^0-\bar{B}^0$ mixing \cite{Aliev4}, $\rho$ parameter
\cite{Soni}, and neutron electric-dipole moment \cite{D.Bowser},
that yields $\bar{\xi}^D_{N,ib}\sim 0$ and $\bar{\xi}^D_{N,ij}\sim
0$, where the indices $i$, $j$ denote d and s quarks, and
$\bar{\xi}^U_{N,tc}<<\bar{\xi}^U_{N,tt}$. Therefore, we take into
account only the Yukawa couplings of b and t quarks, $
\bar{\xi}^U_{N,tt}$, $\bar{\xi}^D_{N,bb}$ and also
$\bar{\xi}^D_{N,\tau\tau}$. There is also a  restriction  on the
 Wilson coefficient $C_7^{eff}$ from the BR of $B\rar X_s \gamma$
in Eq.(\ref{data}) as follows \cite{Aliev4},
\bea 0.257\leq|C_7^{eff}|\leq 0.439. \eea

In the following subsections, we calculate the dependencies  of the
differential branching ratio $(dBR/ds)$ and $A_{FB}(s)$ of the above decays on
the invariant dilepton mass $s$, and also dependencies  of the BR and $A_{FB}$
on the Model III parameters. The results are presented by a series of
graphs, which are plotted for $\ell=\tau$ and  for two different cases of the ratio 
$|r_{tb}|\equiv\Bigg|\frac{\bar{\xi}^D_{N,tt}}{\bar{\xi}^D_{N,bb}}\Bigg|$
where $|r_{tb}|<1$ or $r_{tb}>1$.

\subsection{Numerical results of the exclusive \Bpll decay}
In our numerical calculations for \Bpll decay, we use the results
of the light-cone constituent quark model \cite{Melikhov,Melikhov2}, which can be found from the following expression
\bea f(q^2)=\frac{f(0)}{(1-q^2/T_f)^{n_f}}\, , \eea where the
parameters $f(0)$, $T_f$ and $n_f$ are listed in table \ref{tabpi}.

\FIGURE{\epsfig{file=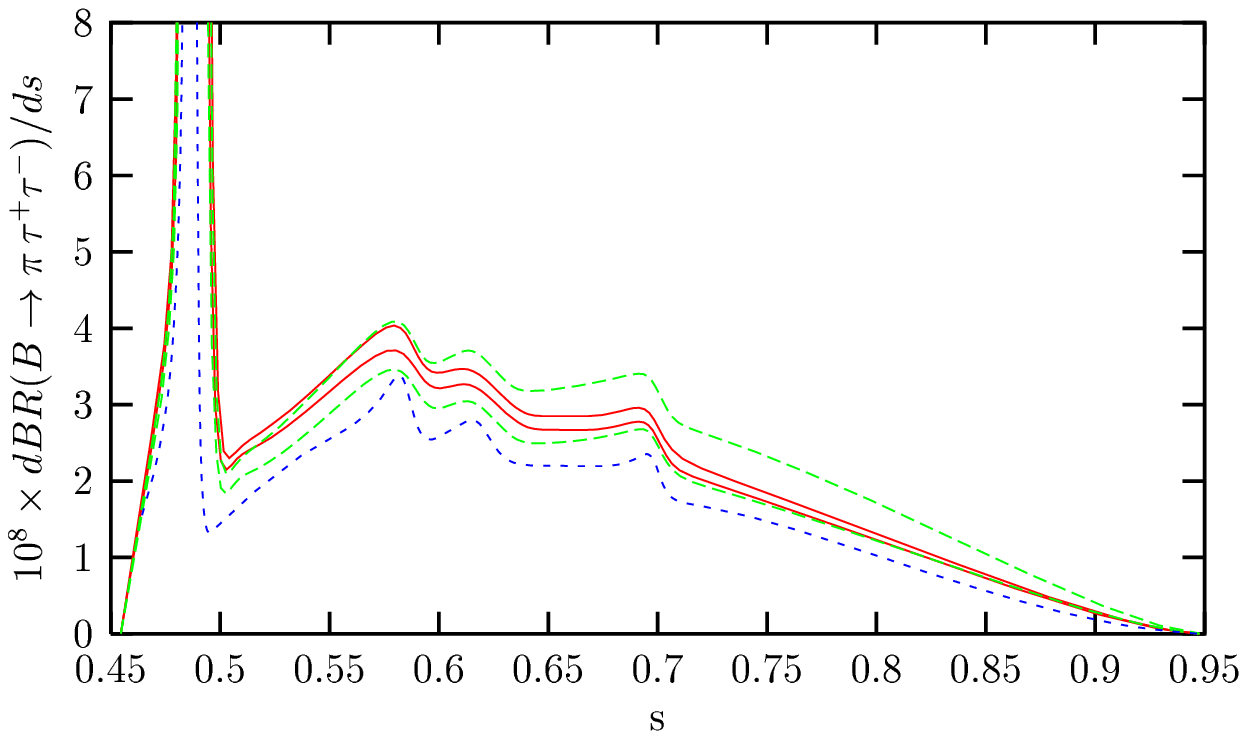,width=9cm}\caption{$dBR/ds$
for \Bptt as a function of $s$ for $\bar{\xi}_{N,bb}^{D}=40\,
m_b$ and $\bar{\xi}_{N,\tau\tau}^{D}=10\, m_{\tau}$, in case that
the ratio $ |r_{tb}| <1$. Here the region
between the solid  curves represents the $dBR/ds$ in Model
III without the NHB effects, while the one between the dashed 
curves is for the $dBR/ds$ with NHB contributions. The SM prediction
is represented by the small dashed  curve }\label{PidGNHBLDa}}
\FIGURE{\epsfig{file=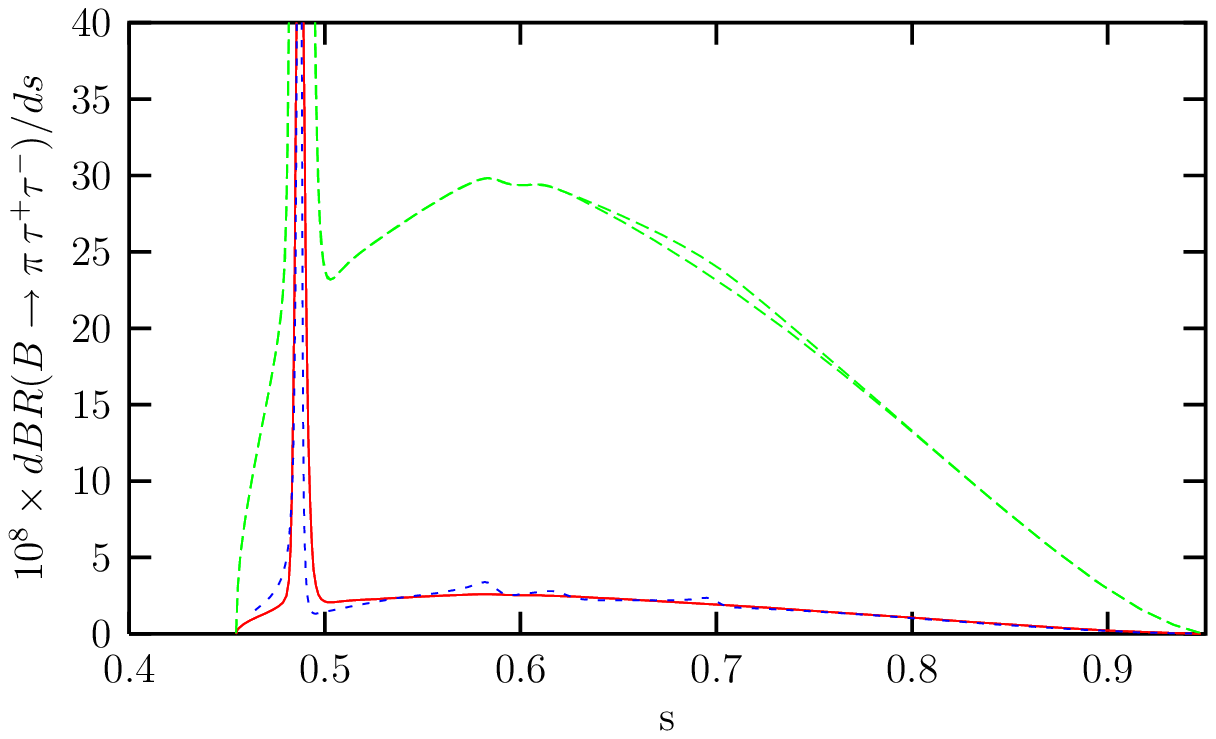,width=9cm}\caption{The same
as Fig.(\ref{PidGNHBLDa}), but for $ r_{tb} >1$ with
$\bar{\xi}_{N,bb}^{D}=0.1\, m_b$ and $\bar{\xi}_{N,\tau\tau}^{D}=
m_{\tau}$.}\label{PidGNHBLDb}}
\FIGURE{\epsfig{file=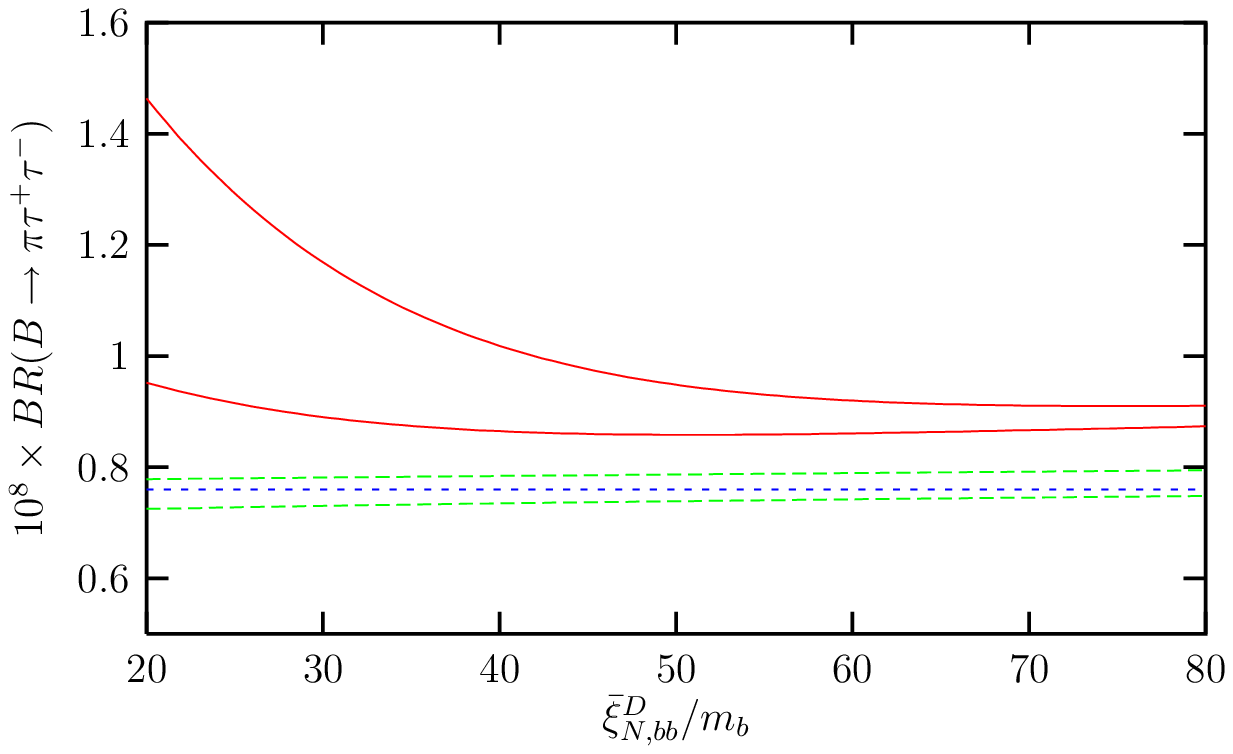,width=9cm}\caption{$BR$ of
\Bptt as a function of $\bar{\xi}_{N,bb}^{D}/m_{b}$  for
$\bar{\xi}_{N,\tau \tau}^{D}= 10 \, m_{\tau}$ and $ |r_{tb}| <1$.
Here $BR$ is restricted in the region between solid (dashed)
curves for $C^{eff}_7 >0$ ($C^{eff}_7 <0$). Small dashed straight
line represents the SM prediction.}\label{PiBRksibba}}
\FIGURE{\epsfig{file=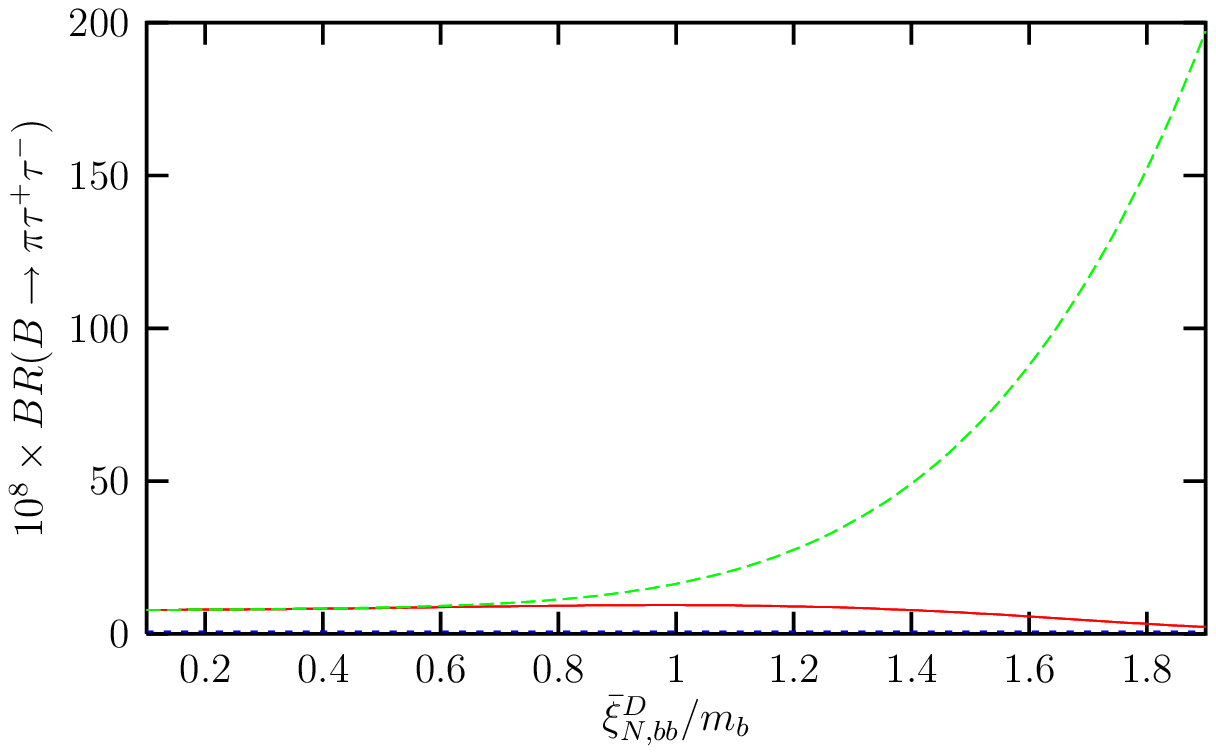,width=9cm}\caption{The same as
Fig.(\ref{PiBRksibba}), but for $ r_{tb} >1$ with
$\bar{\xi}_{N,\tau\tau}^{D}= m_{\tau}$.}\label{PiBRksibbb}}
\FIGURE{\epsfig{file=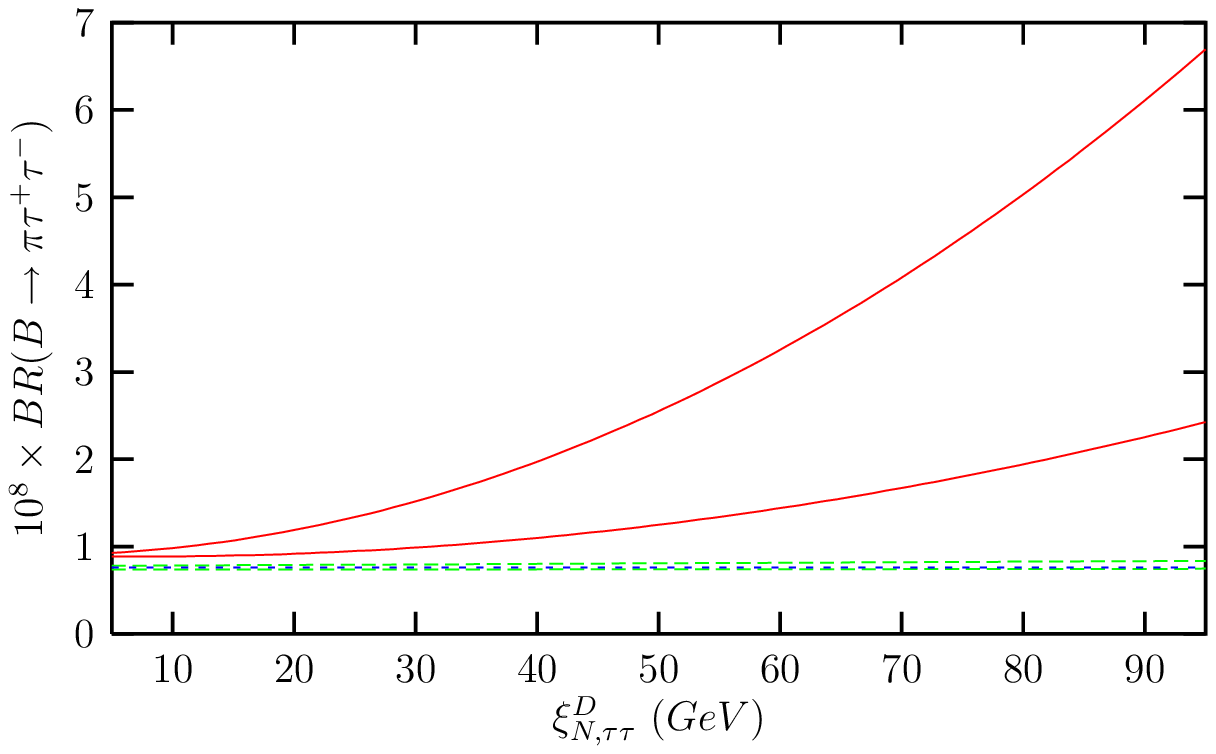,width=9cm}\caption{$BR$ of
\Bptt as a function of  $\bar{\xi}_{N,\tau\tau}^{D}$  for
$\bar{\xi}_{N,bb}^{D}= 40 \, m_{b}$ and $ |r_{tb}| <1$. Here $BR$
is restricted in the region between solid (dashed) curves for
$C^{eff}_7 >0$ ($C^{eff}_7 <0$). Small dashed straight line
represents the SM prediction.}\label{PiBRksitta}}

\FIGURE{\epsfig{file=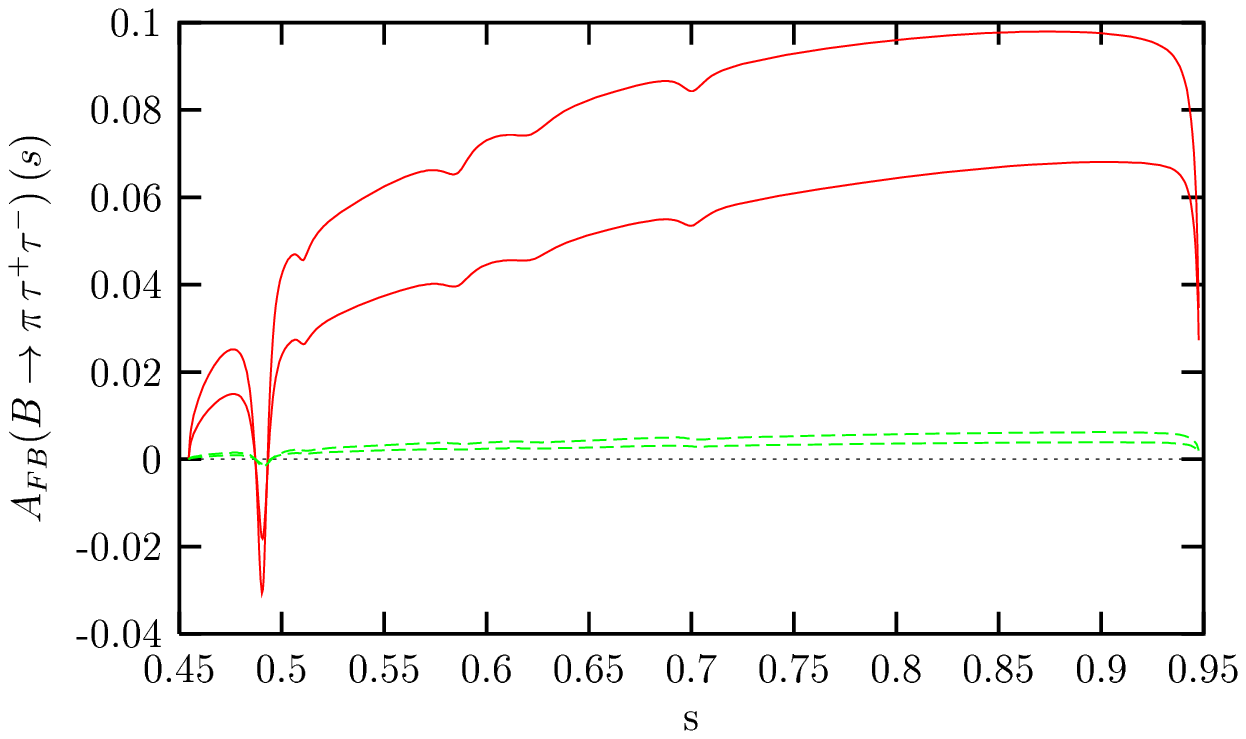,width=9cm}\caption{Differential
$A_{FB}$ for \Bptt as a function of $s$ for
$\bar{\xi}_{N,bb}^{D}=40\, m_b$ and
$\bar{\xi}_{N,\tau\tau}^{D}=10\, m_{\tau}$, in case of the ratio $
|r_{tb}| <1$. Here differential
$A_{FB}$ is restricted in the region between solid (dashed) curves for
$C^{eff}_7 >0$ ($C^{eff}_7 <0$).}\label{PidAFBNHBLDa}}
\FIGURE{\epsfig{file=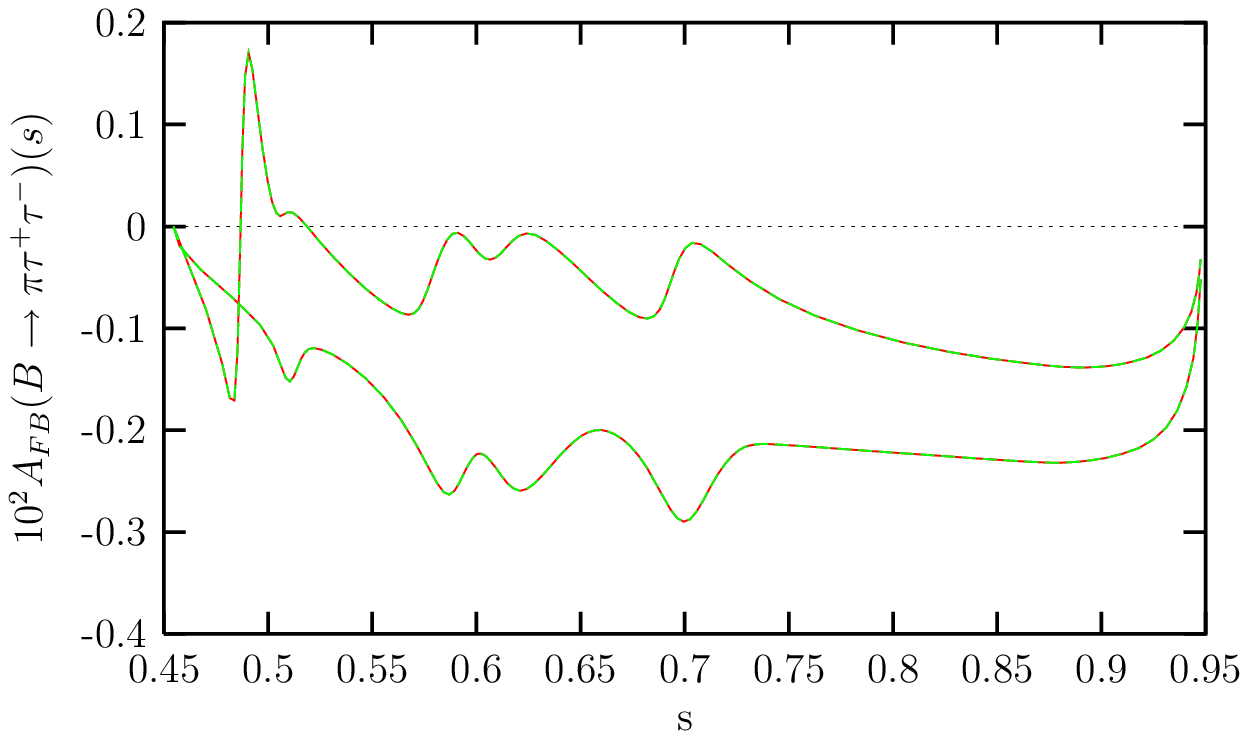,width=9cm}\caption{The same as
Fig.(\ref{PidAFBNHBLDa}), but for $ r_{tb} >1$ with
$\bar{\xi}_{N,bb}^{D}=0.1\, m_b$ and $\bar{\xi}_{N,\tau\tau}^{D}=
m_{\tau}$.}\label{PidAFBNHBLDb}}
\FIGURE{\epsfig{file=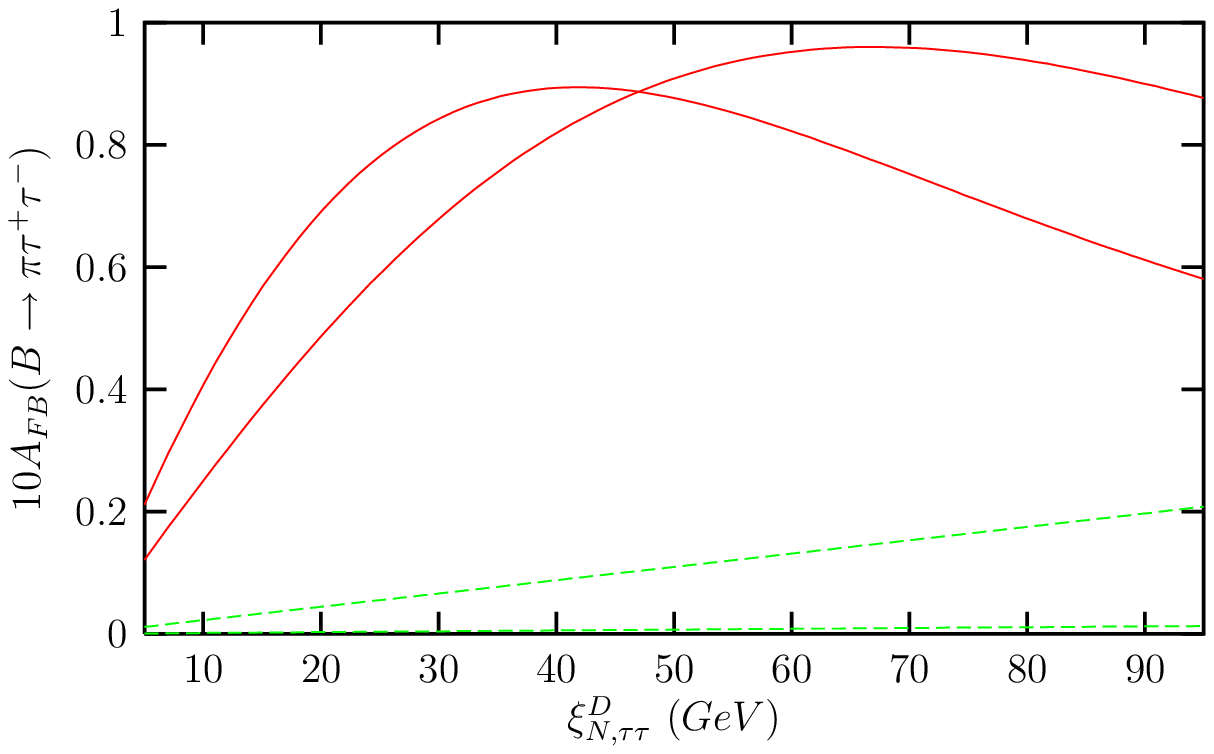,width=9cm}\caption{$A_{FB}$
for \Bptt as a function of  $\bar{\xi}_{N,\tau\tau}^{D}$  for
$\bar{\xi}_{N,bb}^{D}= 40 \, m_{b}$ and $ |r_{tb}| <1$. Here
$A_{FB}$ is restricted in the region between solid (dashed) curves
for $C^{eff}_7 >0$ ($C^{eff}_7 <0$).}\label{PiAFBNHBLDa}}
\FIGURE{\epsfig{file=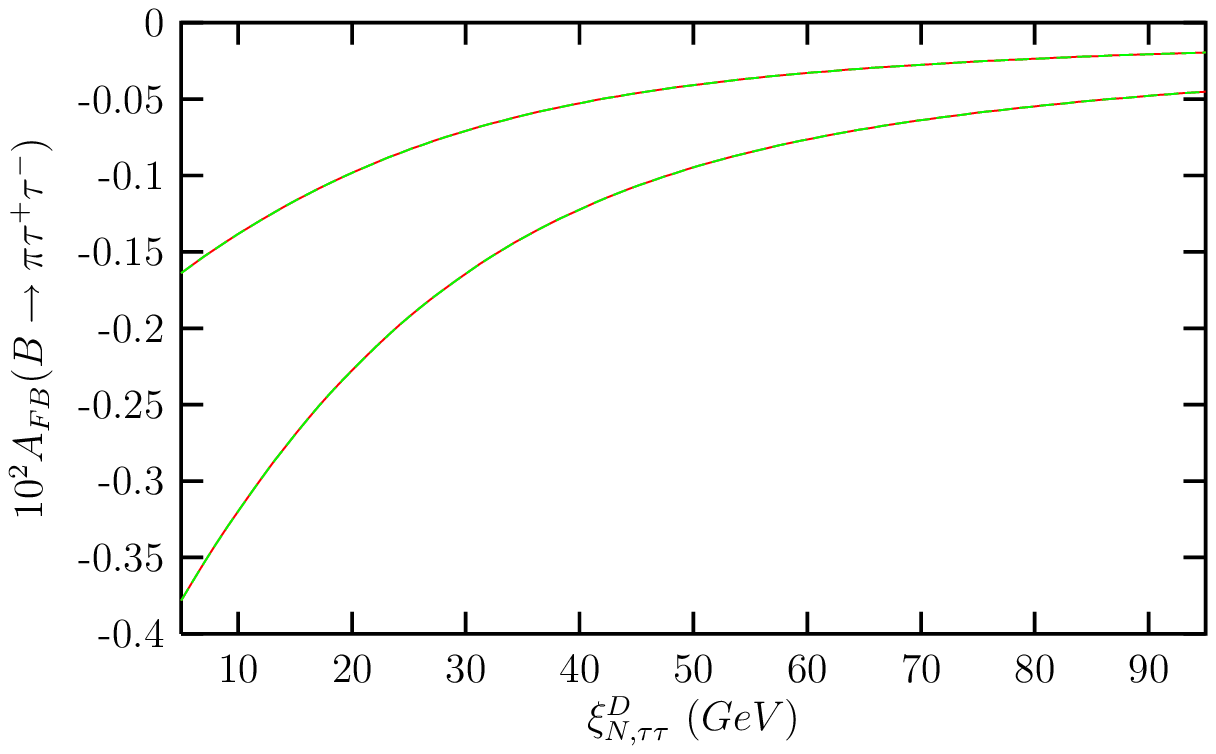,width=9cm}\caption{The same
as Fig.(\ref{PiAFBNHBLDa}), but for $r_{tb}>1$ with
$\bar{\xi}_{N,bb}^{D}= 0.1\, m_b$.}\label{PiAFBNHBLDb}}

In Fig. (\ref{PidGNHBLDa}), we plot the dependence of the
$dBR/ds$ on the invariant dilepton mass $s$, for
$|r_{tb}|<1$ and $C^{eff}_7 >0$ cases, by taking into account 
the long distance effects. Here the region
between the solid (red) curves represents the $dBR/ds$ in Model
III without the NHB effects, while the one between the dashed
(green) curves is for the $dBR/ds$ with NHB contributions. In both
cases, we observe an enhancement compared to the SM prediction,
which is represented by the small dashed (blue) curve. This enhancement
reaches  up to  $70\,\%$ and $50\,\%$ as compared with the SM and
the Model III prediction without NHB effects, respectively. In
Fig. (\ref{PidGNHBLDb}), the same comparison is made for
$r_{tb}>1$ case and we see that the contributions coming from
the NHB effects are extremely large. These figures explicitly show the
size of the NHB effects on the exclusive \Bptt decay.

From now on in figures we plot, the regions bounded by the {\bf
solid} (red) curves represent the $C_7^{eff}>0$ case and the regions
bounded by the {\bf dashed} (green) curves represent the
$C_7^{eff}<0$ case while the {\bf small dashed} (blue) curves are
for the SM predictions for the relevant observable.

\TABULAR{|c| c c |}
 {\hline\hline
   %after \\: \hline or \cline{col1-col2} \cline{col3-col4} ...
   &~~~~~~~~~~~~~~~~~~~~~~$ BR(B\rightarrow\pi\tau^+\tau^-)$&\\\hline
   $(\rho;\eta)$& $|V_{cb}|=0.037$ & $|V_{cb}|=0.043$  \\\hline
  $(0.3;0.34)$ & $0.69\times 10^{-8}$  & $0.93\times 10^{-8}$  \\
  $(-0.3;0.34)$ & $0.60\times 10^{-8}$ &$0.81\times 10^{-8}$  \\
  $(-0.07;0.34)$ & $0.62\times 10^{-8}$ & $0.83\times 10^{-8}$ \\
  \hline\hline}
 {\label{tabpibr}The values of the total branching ratio for \Bptt decay in the SM,
 at three different sets of the Wolfenstein parameters $(\rho;\eta)$.}

We present the dependence of the BR on the parameter
$\bar{\xi}_{N,bb}^{D}/m_{b}$ in
Figs.(\ref{PiBRksibba}-\ref{PiBRksibbb}), where the first one is
for $|r_{tb}|<1$ and the latter for $r_{tb}>1$. Our prediction for
the BR of the \Bptt decay in the SM including the long distance
effects is
 \bea BR(B\rightarrow \pi \tau^+ \tau^-)=0.76\times
10^{-8}. \eea
We also give the SM  values of the total branching ratio for \Bptt decay 
at three different sets of the Wolfenstein parameters $(\rho;\eta)$ in table \ref{tabpibr}.
As seen from Fig. (\ref{PiBRksibba}), for $|r_{tb}|<1$, the $C_7^{eff}<0$ case 
almost coincides with
the SM prediction. However, for
$C_7^{eff}>0$, we observe  an enhancement which is  $1.5-2$ times of
the SM prediction; but this enhancement decreases with the
increasing values of the $\bar{\xi}_{N,bb}^{D}/m_{b}$ parameter.
In case of  $r_{tb}>1$ (Fig. (\ref{PiBRksibbb})),  extremely large enhancement, $2-3$ orders 
larger compared  the SM case, is reached for $C_7^{eff}<0$ case.

We plot the dependence of the BR on the parameter
$\bar{\xi}_{N,\tau \tau}^{D}$ in Fig. (\ref{PiBRksitta})   for
$|r_{tb}|<1$ . From this figure, we again observe an enhancement
as in the $\bar{\xi}_{N,bb}^{D}/m_{b}$ dependence and this is the contribution due to the NHB effects. However,  the
behavior of this dependence is opposite to that of  $\bar{\xi}_{N,bb}^{D}/m_{b}$ dependence: the BR increases with
the increasing values of the $\bar{\xi}_{N,\tau \tau}^{D}$. The SM
prediction again lies in the region bounded by the $C_7^{eff}<0$
case.

The dependence of the differential $A_{FB}$  on the invariant
dilepton mass $s$ for the \Bptt decay is presented in
Fig. (\ref{PidAFBNHBLDa}) (Fig. (\ref{PidAFBNHBLDb})) for
$|r_{tb}|<1$ ($r_{tb}>1$) case. Since $A_{FB}$ arises in the 2HDM
only when the NHB effects are taken into account, it provides a
good probe to test  these effects. For $|r_{tb}|<1$, although
 $A_{FB}(s)$ is very small for $C_7^{eff}<0$ case, it is
considerably enhanced  for $C_7^{eff}>0$ case. For $r_{tb}>1$,
$A_{FB}(s)$  for $C_7^{eff}<0$ and $C_7^{eff}>0$ cases completely
coincide and its magnitude is one order smaller than the $|A_{FB}(s)|$
for $|r_{tb}|<1$ case.

Figs.(\ref{PiAFBNHBLDa}) and (\ref{PiAFBNHBLDb}) are devoted to the
$\bar{\xi}_{N,\tau \tau}^{D}$ dependence of the $A_{FB}$ of the
\Bptt decay for $|r_{tb}|<1$ and $r_{tb}>1$ cases, respectively.  
As can be observed from Fig.(\ref{PiAFBNHBLDa}), $A_{FB}$  is quite 
sensitive to the parameter $\bar{\xi}_{N,\tau \tau}^{D}$ especially for
$C_7^{eff}>0$. It can reach 10\% for $\bar{\xi}_{N,\tau \tau}^{D}\sim 60$.
For $r_{tb}>1$ case shown in
Fig.(\ref{PiAFBNHBLDb}), the $C_7^{eff}<0$ and $C_7^{eff}>0$ cases
completely coincide and $|A_{FB}|$ decreases
with the increasing values of the $\bar{\xi}_{N,\tau \tau}^{D}$. In addition,  
its value  is one order smaller than the $|r_{tb}|<1$ case.

\subsection{Numerical results of the exclusive \Brll decay}
In our numerical calculation for \Brll decay, we use three parameter fit
of the light-cone QCD sum rule \cite{Ball} which can be written in the following form
\bea\label{formrho}
F(q^2)=\frac{F(0)}{1-a_F~q^2/m_B^2+b_F(q^2/m_B^2)^2} \eea where
the values of the parameters $F(0)$, $a_F$ and $b_F$ are given in
table (\ref{tabrho}). The form factors $A_0$ and $A_3$ can be
found from the following parametrization,
\bea\label{paramet} A_0&=&A_3-\frac{T_3~q^2}{m_\rho
m_b},\nonumber\\
A_3&=&\frac{m_B+m_\rho}{2m_\rho}A_1-\frac{m_B-m_\rho}{2m_\rho}A_2.
 \eea

\FIGURE{\epsfig{file=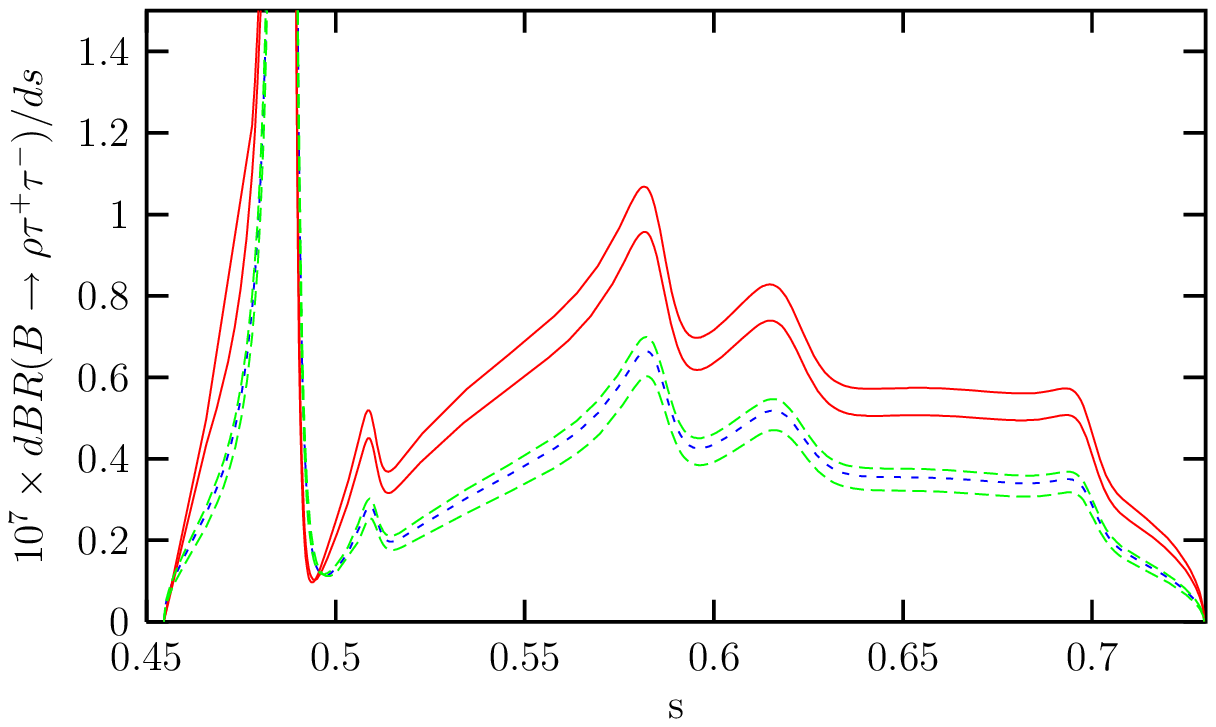,width=9cm}\caption{$dBR/ds$
 for \Brtt as a function of $s$ for $\bar{\xi}_{N,bb}^{D}=40\,
m_b$ and $\bar{\xi}_{N,\tau\tau}^{D}=10\, m_{\tau}$, in case of
the ratio $ |r_{tb}| <1$. Here $dBR/ds$ is
restricted in the region between solid (dashed) curves for
$C^{eff}_7 >0$ ($C^{eff}_7 <0$. Small dashed curve
represents the SM prediction.}\label{dGNHBLDa}}
\FIGURE{\epsfig{file=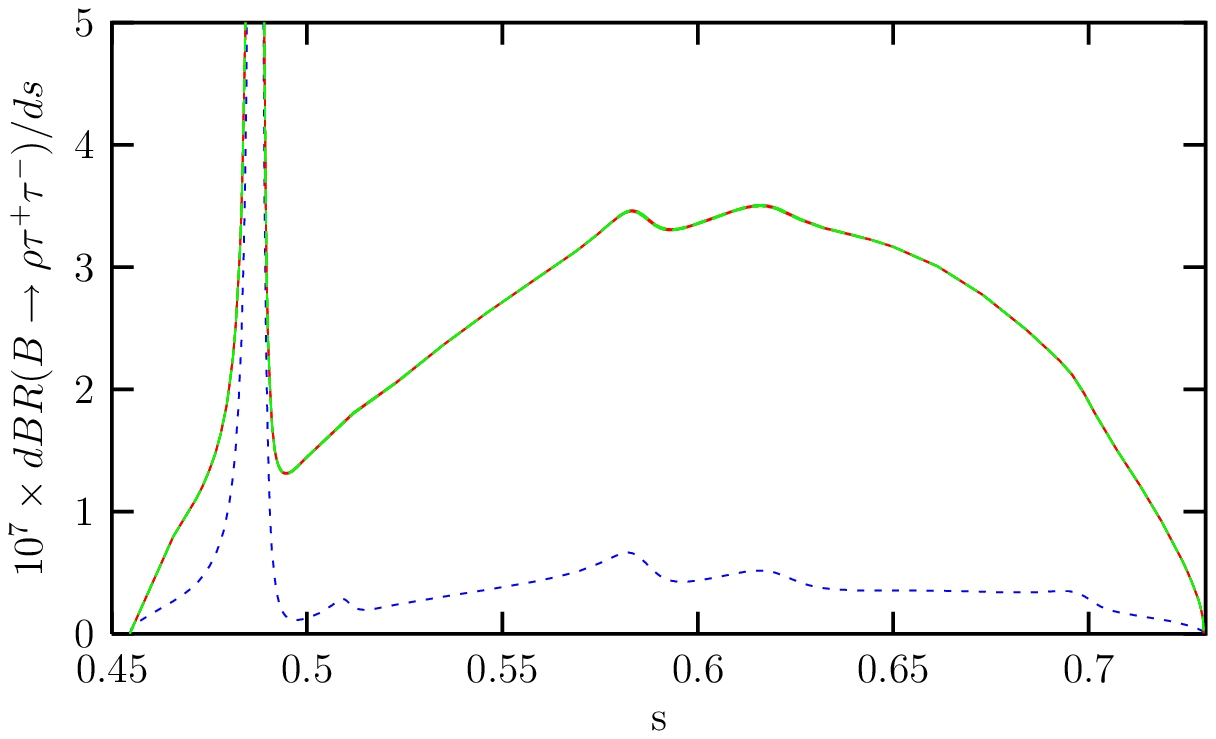,width=9cm}\caption{The same
as Fig.(\ref{dGNHBLDa}), but for $ r_{tb} >1$ with
$\bar{\xi}_{N,bb}^{D}=0.1\, m_b$ and $\bar{\xi}_{N,\tau\tau}^{D}=
m_{\tau}$.}\label{dGNHBLDb}}
\FIGURE{\epsfig{file=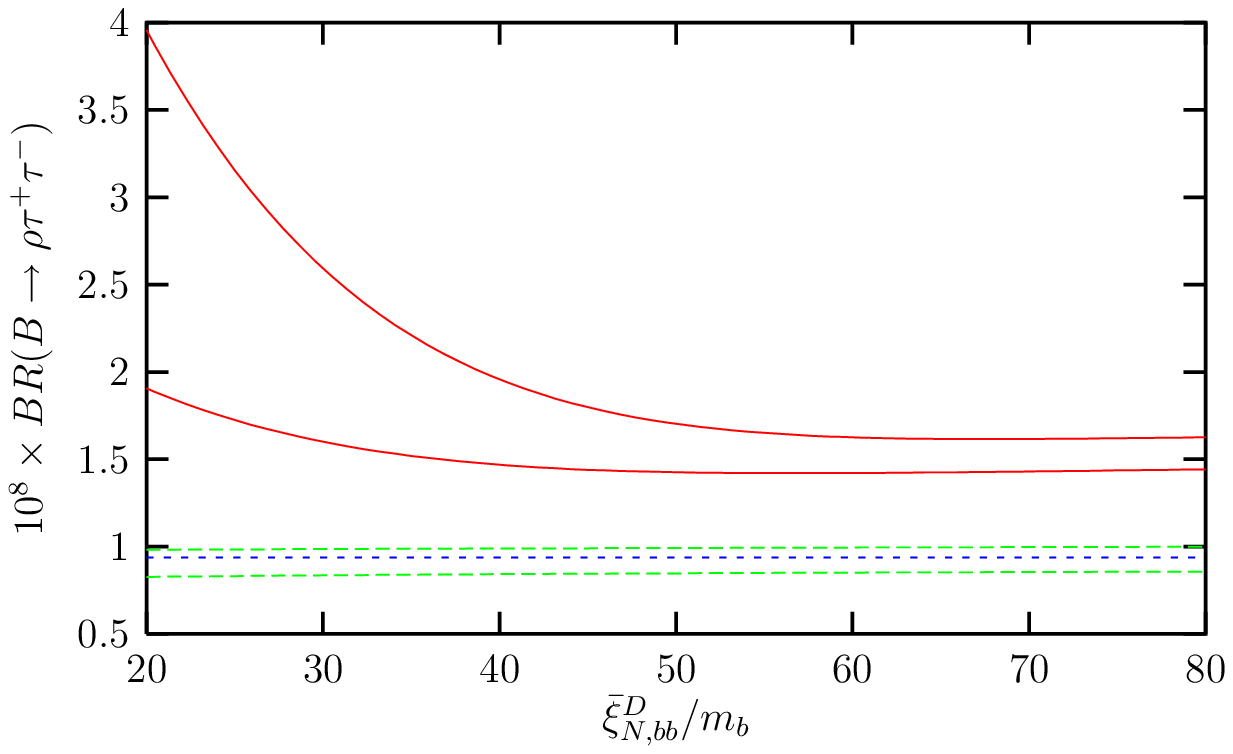,width=9cm}\caption{$BR$ for \Brtt
as a function of  $\bar{\xi}_{N,bb}^{D}/m_{b}$  for $\bar{\xi}_{N,\tau
\tau}^{D}= 10 \, m_{\tau}$ and $ |r_{tb}| <1$. Here $BR$ is
restricted in the region between solid (dashed) curves for
$C^{eff}_7 >0$ ($C^{eff}_7 <0$). Small dashed straight line 
represents the SM prediction.}\label{BRksibba}}
\FIGURE{\epsfig{file=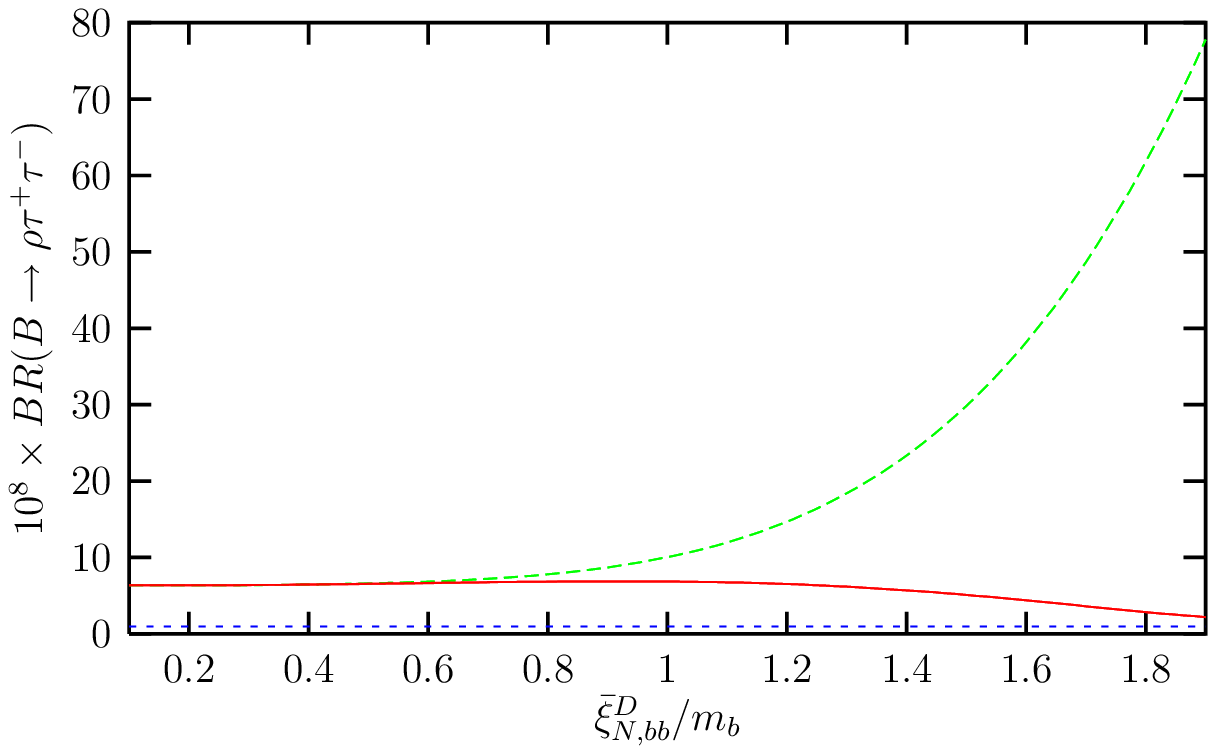,width=9cm}\caption{The same
as Fig.(\ref{BRksibba}), but for $ r_{tb} >1$ with
$\bar{\xi}_{N,\tau\tau}^{D}= m_{\tau}$.}\label{BRksibbb}}
\FIGURE{\epsfig{file=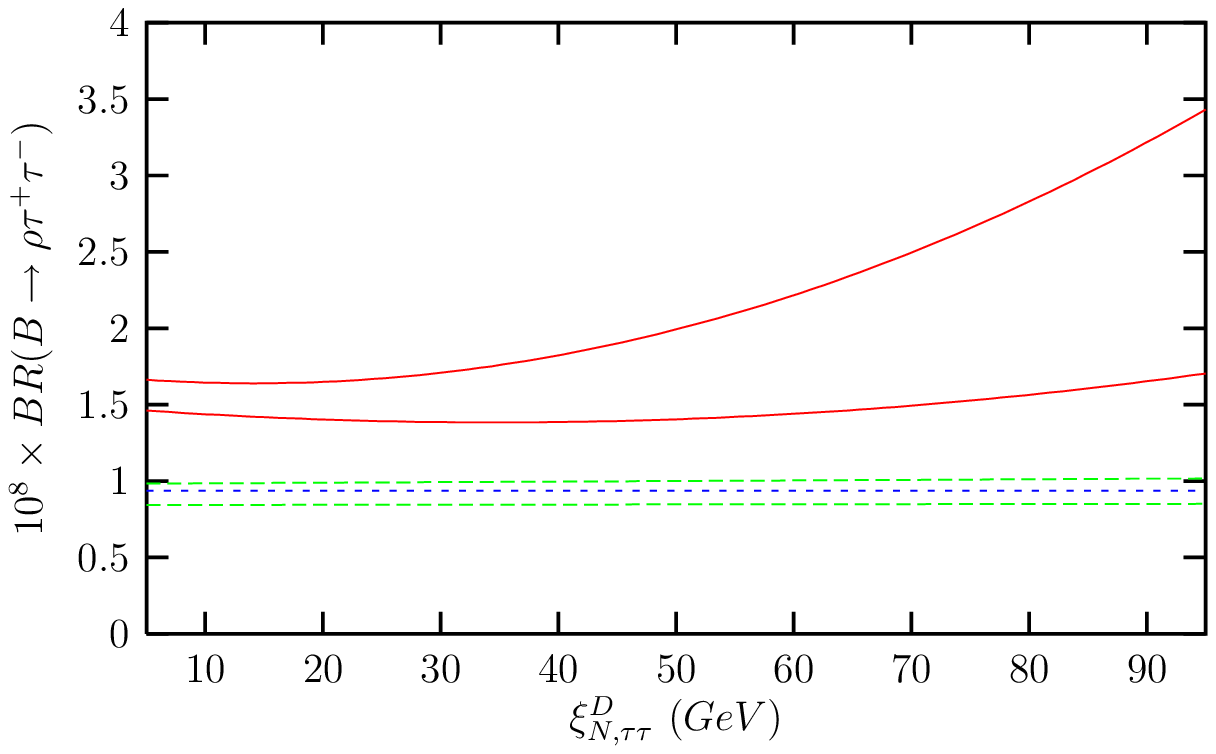,width=9cm}\caption{$BR$ for \Brtt as a
function of  $\bar{\xi}_{N,\tau\tau}^{D}$  for
$\bar{\xi}_{N,bb}^{D}= 40 \, m_{b}$ and $ |r_{tb}| <1$. Here $BR$
is restricted in the region between solid (dashed) curves for
$C^{eff}_7 >0$ ($C^{eff}_7 <0$). Small dashed straight line
represents the SM prediction.}\label{BRksitta}}

\FIGURE{\epsfig{file=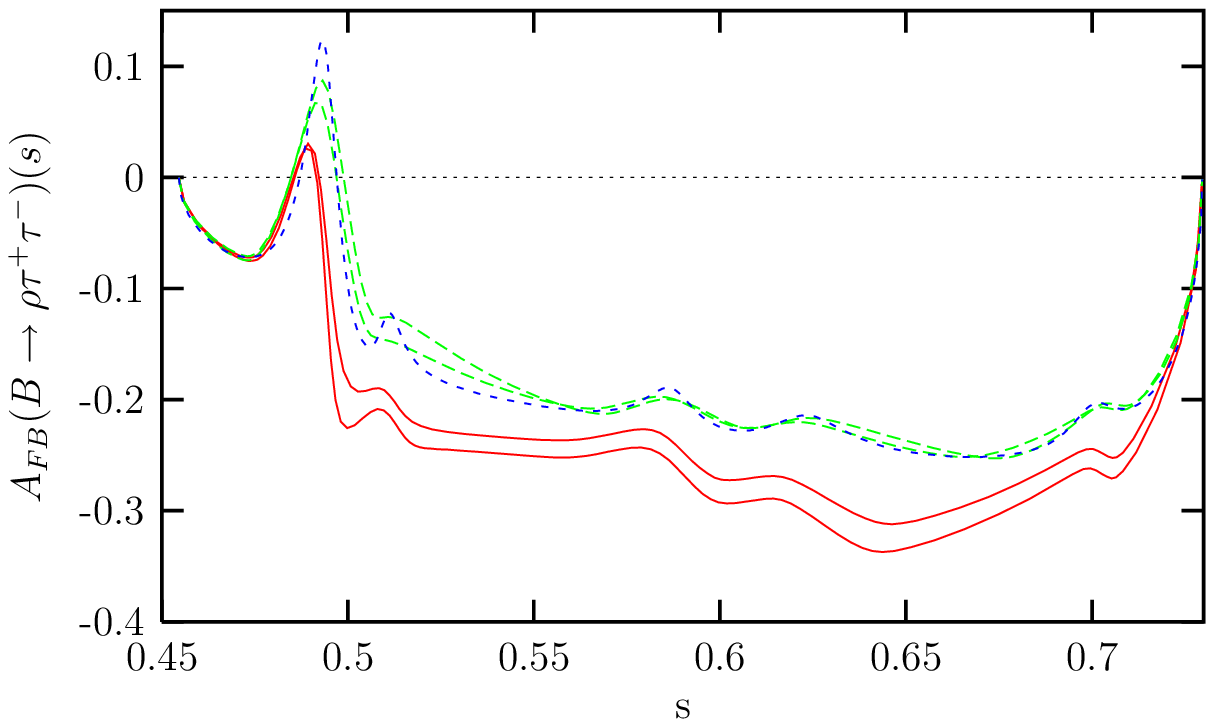,width=9cm}\caption{ 
$A_{FB}(s)$ for \Brtt as a function of $s$ for
$\bar{\xi}_{N,bb}^{D}=40\, m_b$ and
$\bar{\xi}_{N,\tau\tau}^{D}=10\, m_{\tau}$, in case of the ratio $
|r_{tb}| <1$. Here $A_{FB}(s)$ is
restricted in the region between solid (dashed) curves for
$C^{eff}_7 >0$ ($C^{eff}_7 <0$). Small dashed curve 
represents the SM prediction.}\label{dAFBNHBLDa}}
\FIGURE{\epsfig{file=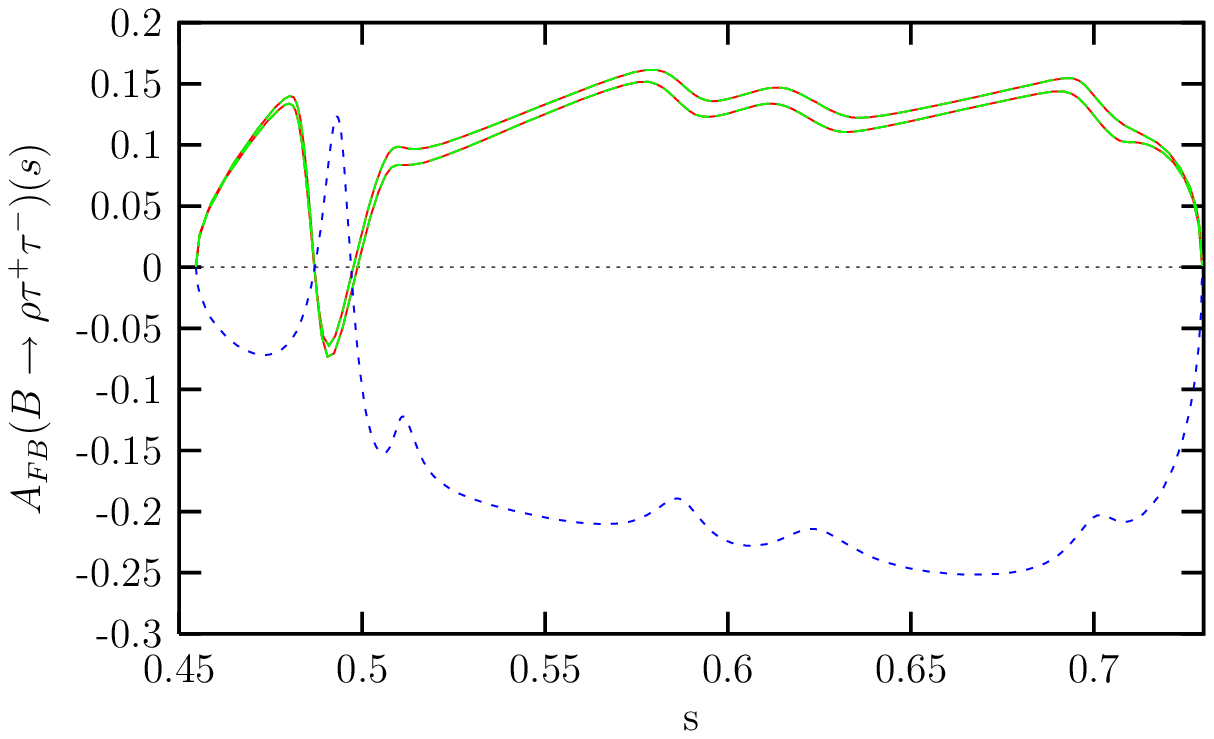,width=9cm}\caption{The same
as Fig.(\ref{dAFBNHBLDa}), but for $ r_{tb} >1$ with
$\bar{\xi}_{N,bb}^{D}=0.1\, m_b$ and $\bar{\xi}_{N,\tau\tau}^{D}=
m_{\tau}$.}\label{dAFBNHBLDb}}
\FIGURE{\epsfig{file=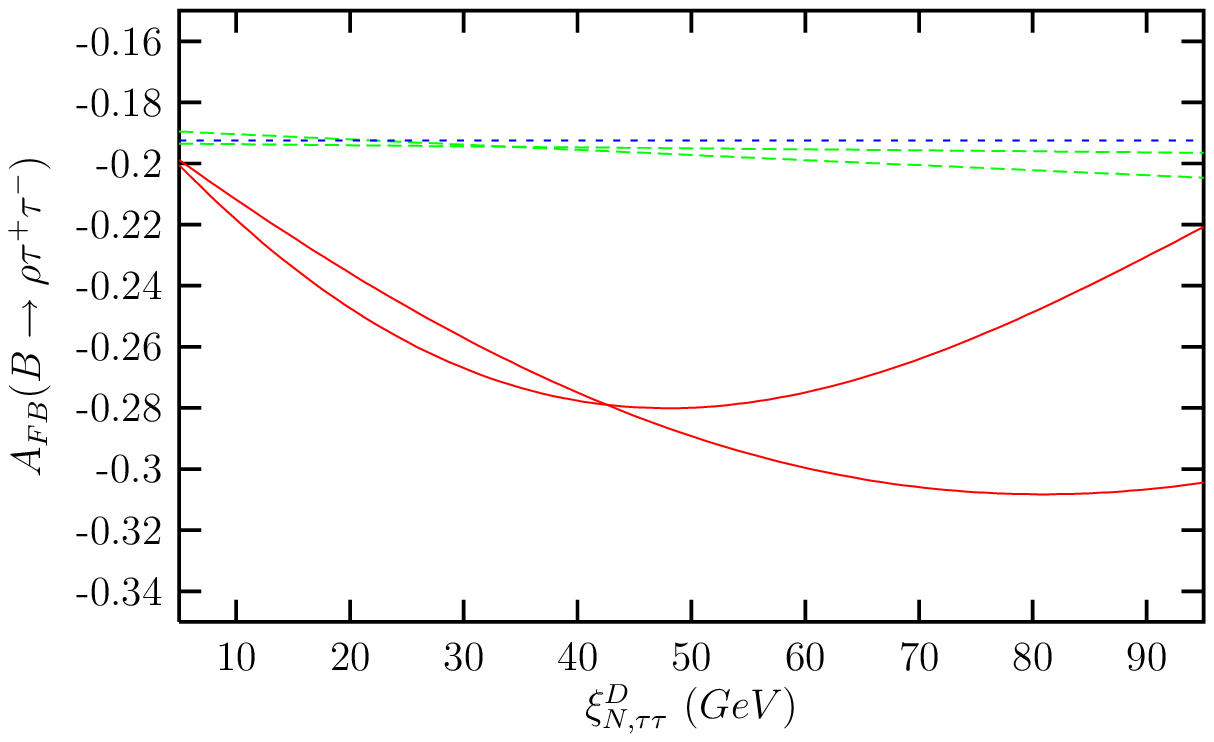,width=9cm}\caption{$A_{FB}$
for \Brtt as a function of  $\bar{\xi}_{N,\tau\tau}^{D}$  for
$\bar{\xi}_{N,bb}^{D}= 40 \, m_{b}$ and $ |r_{tb}| <1$. Here
$A_{FB}$ is restricted in the region between solid (dashed) curves
for $C^{eff}_7 >0$ ($C^{eff}_7 <0$). Small dashed straight line
represents the SM prediction.}\label{AFBNHBLDa}}
\FIGURE{\epsfig{file=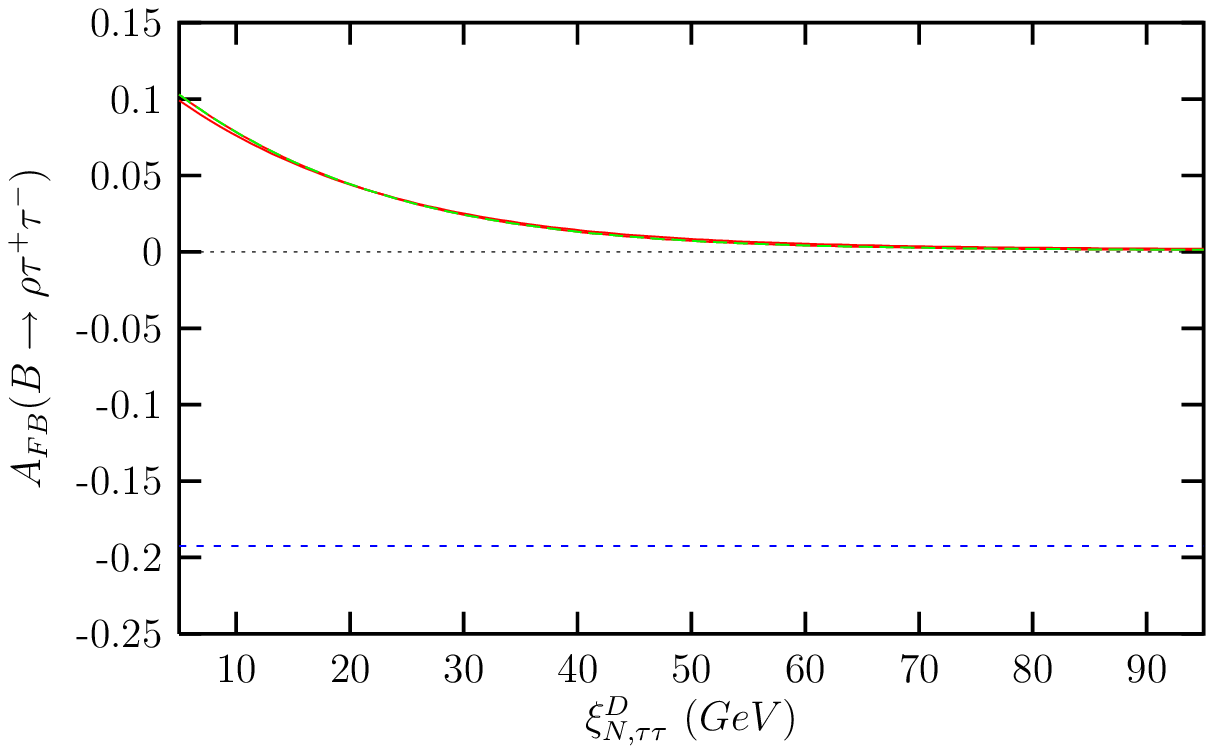,width=9cm}\caption{The same
as Fig.(\ref{AFBNHBLDa}), but for $r_{tb}>1$ with
$\bar{\xi}_{N,bb}^{D}= 0.1\, m_b$.}\label{AFBNHBLDb}}

We first consider the dependence of $dBR/ds$ on the invariant dilepton mass $s$ for the \Brtt decay. This is plotted in Fig.(\ref{dGNHBLDa}) for
$\bar{\xi}^D_{N,bb}=40~m_b$ and
$\bar{\xi}^D_{N,\tau\tau}=10~m_\tau$, in case of the ratio
$|r_{tb}|<1$ by taking into account the long distance effects.
We conclude from this graph that, $dBR/ds$ almost coincides with the SM result for $C_7^{eff}<0$ case, while for $C_7^{eff}>0$ it is considerably enhanced.
As for the $r_{tb}>1$ case (Fig. (\ref{dGNHBLDb}))where we take
$\bar{\xi}^D_{N,bb}=0.1~m_b$ and $\bar{\xi}^D_{N,\tau\tau}=m_\tau$, we observe an enhancement of one order as compared with the $|r_{tb}|<1$ and also the SM cases. Here $C_7^{eff}>0$ and $C_7^{eff}<0$ cases completely coincide.
 \TABULAR{|c| c c |c|}
 {\hline\hline
   %after \\: \hline or \cline{col1-col2} \cline{col3-col4} ...
   &~~~~~~~~~~~~~~~~~~$ BR(B\rightarrow\rho\tau^+\tau^-)$& &$A_{FB}(B\rightarrow\rho\tau^+\tau^-)$\\\hline
   $(\rho;\eta)$& $|V_{cb}|=0.037$ & $|V_{cb}|=0.043$&  \\\hline
  $(0.3;0.34)$ & $0.94\times 10^{-8}$  & $1.26\times 10^{-8}$&-0.17  \\
  $(-0.3;0.34)$ & $0.72\times 10^{-8}$ &$0.97\times 10^{-8}$ &-0.20 \\
  $(-0.07;0.34)$ & $0.76\times 10^{-8}$ & $1.02\times 10^{-8}$&-0.19 \\
  \hline\hline}
 {\label{tabrhobr}The values of the total branching ratio and the forward-backward 
 asymmetry for \Brtt decay in the SM,
 at three different sets of the Wolfenstein parameters $(\rho;\eta)$.}

The dependence of the BR on one of the free  parameters of the model III,
$\bar{\xi}_{N,bb}^{D}/m_{b}$, is given in Figs.
(\ref{BRksibba}) and (\ref{BRksibbb}) for $|r_{tb}|<1$
and $r_{tb}>1$, respectively. Our prediction for the BR of the
\Brtt decay in the SM including the long distance effects is
\bea
BR(B\rightarrow \rho \tau^+ \tau^-)=0.94\times 10^{-8}. \eea
We also give the SM  values of the total branching ratio together with $A_{FB}$ values
 for \Brtt decay  at three different sets of the Wolfenstein parameters $(\rho;\eta)$ 
 in table \ref{tabrhobr}. As seen from Fig. (\ref{BRksibba}),
for $|r_{tb}|<1$ where we take
$\bar{\xi}_{N,\tau \tau}^{D}= 10 \, m_{\tau}$, the $C_7^{eff}<0$ case 
coincides with the SM prediction. When $C_7^{eff}>0$, however, the
BR is enhanced by $2.5-5$ times of the SM prediction; but this
enhancement decreases with the increasing values of the
$\bar{\xi}_{N,bb}^{D}/m_{b}$ parameter. For $r_{tb}>1$, we take
$\bar{\xi}_{N,\tau \tau}^{D}= \, m_{\tau}$ and observe an
enhancement for both $C_7^{eff}<0$ and $C_7^{eff}>0$ cases. For
$C_7^{eff}<0$, it is  one order larger than the SM value while for $C_7^{eff}>0$,
 the order of enhancement is the same as that in $|r_{tb}|<1$ case.

We plot the dependence of the BR on $\bar{\xi}_{N,\tau \tau}^{D}$, the other free parameter of model III, in Fig.
(\ref{BRksitta}) for $|r_{tb}|<1$. Here, we take $\bar{\xi}_{N,bb}^{D}= 40 \, m_{b}$ 
 and  see that  the BR is not sensitive to $\bar{\xi}_{N,\tau \tau}^{D}$ for $C_7^{eff}<0$ and it is almost the same as the the SM prediction. However for  $C_7^{eff}>0$, the BR is quite sensitive to $\bar{\xi}_{N,\tau \tau}^{D}$ and increases as it increases.

The dependence of $A_{FB}(s)$ of the \Brtt decay on
the invariant dilepton mass $s$ is represented in
Fig.(\ref{dAFBNHBLDa}) (Fig.(\ref{dAFBNHBLDb})) for $|r_{tb}|<1$
($r_{tb}>1$) case. For $|r_{tb}|<1$, there is an enhancement on
$|A_{FB}(s)|$ for $C_7^{eff}>0$, while for $C_7^{eff}<0$ it is almost the same as the SM prediction. For $r_{tb}>1$, all Model III
predictions for $A_{FB}(s)$ almost coincide but with
a flip in the sign as compared to the SM prediction.

Finally we present the dependence of the $A_{FB}$ on the
$\bar{\xi}_{N,\tau \tau}^{D}$ parameter. Our prediction for the
$A_{FB}$ of the \Brtt decay in the SM is
\bea A_{FB}(B\rar
\rho\,\tau^+ \tau^-)=\,-0.193 \eea
(See also table \ref{tabrhobr}.)
In Fig. (\ref{AFBNHBLDa}), this
dependence is plotted for the ratio $|r_{tb}|<1$ and
$\bar{\xi}_{N,bb}^{D}= 40\, m_b$. Although the $C_7^{eff}<0$ case
almost coincides with the SM value, there is an enhancement for $|A_{FB}|$ up to
the $50 \%$ of the SM value for the moderate values of the
$\bar{\xi}_{N,\tau \tau}^{D}$ parameter, for $C_7^{eff}>0$ case.
For $r_{tb}>1$ case where we take
$\bar{\xi}_{N,bb}^{D}= 0.1\, m_b$, $|A_{FB}|$ can reach at most  half
of the SM value and drops to zero for large values of $\bar{\xi}_{N,\tau \tau}^{D}$. Here, $C_7^{eff}<0$ and
$C_7^{eff}>0$ cases almost coincide.

Finally, we would like to comment briefly about the NHB effects on the CP
violating asymmetry, $A_{CP}$, for \Bpll and \Brll decays. As pointed out before
\cite{Kruger1}-\cite{Bertoluni},  in the SM there is a considerable $A_{CP}$ in the 
partial rates for these decays 
because all three CKM factors contributing are at the same order. Further, the 2HDM 
contributions to $A_{CP}$ have been investigated  in \cite{Aliev1,Iltan1}
and it is shown that
since charged Higgs contributions give rise to constructive interference to the SM
result, $A_{CP}$ decreases while the BR increases  for \Bpll and \Brll decays in the 2HDM. 
We expect that including the NHB effects will further decrease 
magnitude of $A_{CP}$ with respect to its value without NHB effects. To see this, 
consider $A_{CP}$
between $B\rightarrow M \ell^+ \ell^-$ and  $B\rightarrow \bar{M} \ell^+ \ell^-$
decays for $M=\pi , \rho$, which can be written as \cite{Erhanx, GITuran}
\bea
A_{CP} & \sim & \frac{\int~ ds ~ Im(C^{eff}_7) Im(C^{eff}_9) {\cal F}}{\int ~ ds ~ \Delta}
\nnb
\eea
where ${\cal F}$ is a function of various form factors for the decays we consider and
$\Delta$ is proportional to the one given by  Eq.(\ref{deltapi}) and (\ref{deltarho}) 
for $\pi$ and $\rho$,
respectively. As can be seen from the equation above, the numerator of the $A_{CP}$ ratio
is free from the NHB contributions while the denominator gets this additional 
contribution so that magnitude of $A_{CP}$ will decrease with the inclusion of the NHB
contributions.

\subsection{Conclusion}

In this paper we have investigated the physical observables, $BR$
and $A_{FB}$, related to the exclusive \Bpll and \Brll decays in
the general 2HDM including the NHB effects. We have found that NHB
effects are quite sizable, leading to considerable enhancements on
these physical observables. An experimental observation of the
$A_{FB}$ in the \Bpll decay, which is absent in the SM, would be a
very powerful and direct test of the 2HDM and the existence of
NHB. In conclusion we say that the exclusive \Bpll and \Brll
decays provide very useful testing ground for the new physics
beyond the SM.
%\newpage
%\newpage
%{\bf \LARGE {Appendix}} \\
\appendix
\section{The operator basis}\label{App1}
The operator basis in the  2HDM for our process is
\cite{Dai,Grinstein,Misiak}
\begin{eqnarray}
 O_1 &=& (\bar{d}_{L \alpha} \gamma_\mu c_{L \beta})
               (\bar{c}_{L \beta} \gamma^\mu b_{L \alpha}), \nonumber   \\
 O_2 &=& (\bar{d}_{L \alpha} \gamma_\mu c_{L \alpha})
               (\bar{c}_{L \beta} \gamma^\mu b_{L \beta}),  \nonumber   \\
 O_3 &=& (\bar{d}_{L \alpha} \gamma_\mu b_{L \alpha})
               \sum_{q=u,d,s,c,b}
               (\bar{q}_{L \beta} \gamma^\mu q_{L \beta}),  \nonumber   \\
 O_4 &=& (\bar{d}_{L \alpha} \gamma_\mu b_{L \beta})
                \sum_{q=u,d,s,c,b}
               (\bar{q}_{L \beta} \gamma^\mu q_{L \alpha}),   \nonumber  \\
 O_5 &=& (\bar{d}_{L \alpha} \gamma_\mu b_{L \alpha})
               \sum_{q=u,d,s,c,b}
               (\bar{q}_{R \beta} \gamma^\mu q_{R \beta}),   \nonumber  \\
 O_6 &=& (\bar{d}_{L \alpha} \gamma_\mu b_{L \beta})
                \sum_{q=u,d,s,c,b}
               (\bar{q}_{R \beta} \gamma^\mu q_{R \alpha}),  \nonumber   \\
 O_7 &=& \frac{e}{16 \pi^2}
          \bar{d}_{\alpha} \sigma_{\mu \nu} (m_b R + m_s L) b_{\alpha}
                {\cal{F}}^{\mu \nu},                             \nonumber  \\
 O_8 &=& \frac{g}{16 \pi^2}
    \bar{d}_{\alpha} T_{\alpha \beta}^a \sigma_{\mu \nu} (m_b R +
m_s L)
          b_{\beta} {\cal{G}}^{a \mu \nu} \nonumber \,\, , \\
 O_9 &=& \frac{e}{16 \pi^2}
          (\bar{d}_{L \alpha} \gamma_\mu b_{L \alpha})
              (\bar{\ell} \gamma^\mu \ell)  \,\, ,    \nonumber    \\
 O_{10} &=& \frac{e}{16 \pi^2}
          (\bar{d}_{L \alpha} \gamma_\mu b_{L \alpha})
              (\bar{\ell} \gamma^\mu \gamma_{5} \ell)  \,\, ,  \nnb \\
 O^u_1 &=& (\bar{d}_{L \alpha} \gamma_\mu u_{L \beta})
               (\bar{u}_{L \beta} \gamma^\mu b_{L \alpha}), \nonumber   \\
 O^u_2 &=& (\bar{d}_{L \alpha} \gamma_\mu u_{L \alpha})
               (\bar{u}_{L \beta} \gamma^\mu b_{L \beta}),  \nonumber   \\
Q_1&=&   \frac{e^2}{16
\pi^2}(\bar{d}^{\alpha}_{L}\,b^{\alpha}_{R})\,(\bar{\ell}\ell ) \,
, \nnb  \\ Q_2&=&\frac{e^2}{16
\pi^2}(\bar{d}^{\alpha}_{L}\,b^{\alpha}_{R})\, (\bar{\ell}
\gamma_5 \ell ) \, , \nnb \\ Q_3&=&    \frac{g^2}{16
\pi^2}(\bar{d}^{\alpha}_{L}\,b^{\alpha}_{R})\, \sum_{q=u,d,s,c,b
}(\bar{q}^{\beta}_{L} \, q^{\beta}_{R} ) \, ,\nnb \\ Q_4&=&
\frac{g^2}{16 \pi^2}(\bar{d}^{\alpha}_{L}\,b^{\alpha}_{R})\,
\sum_{q=u,d,s,c,b } (\bar{q}^{\beta}_{R} \, q^{\beta}_{L} ) \, ,
\nnb \\ Q_5&=&   \frac{g^2}{16
\pi^2}(\bar{d}^{\alpha}_{L}\,b^{\beta}_{R})\, \sum_{q=u,d,s,c,b }
(\bar{q}^{\beta}_{L} \, q^{\alpha}_{R} ) \, , \nnb \\ Q_6&=&
\frac{g^2}{16 \pi^2}(\bar{d}^{\alpha}_{L}\,b^{\beta}_{R})\,
\sum_{q=u,d,s,c,b } (\bar{q}^{\beta}_{R} \, q^{\alpha}_{L} ) \, ,
\nnb \\ Q_7&=&   \frac{g^2}{16
\pi^2}(\bar{d}^{\alpha}_{L}\,\sigma^{\mu \nu} \, b^{\alpha}_{R})\,
\sum_{q=u,d,s,c,b } (\bar{q}^{\beta}_{L} \, \sigma_{\mu \nu }
q^{\beta}_{R} ) \, , \nnb \\ Q_8&=&    \frac{g^2}{16
\pi^2}(\bar{d}^{\alpha}_{L}\,\sigma^{\mu \nu} \, b^{\alpha}_{R})\,
\sum_{q=u,d,s,c,b } (\bar{q}^{\beta}_{R} \, \sigma_{\mu \nu }
q^{\beta}_{L} ) \, ,  \nnb \\ Q_9&=&   \frac{g^2}{16
\pi^2}(\bar{d}^{\alpha}_{L}\,\sigma^{\mu \nu} \, b^{\beta}_{R})\,
\sum_{q=u,d,s,c,b }(\bar{q}^{\beta}_{L} \, \sigma_{\mu \nu }
q^{\alpha}_{R} ) \, , \nnb \\ Q_{10}&= & \frac{g^2}{16
\pi^2}(\bar{d}^{\alpha}_{L}\,\sigma^{\mu \nu} \, b^{\beta}_{R})\,
\sum_{q=u,d,s,c,b }(\bar{q}^{\beta}_{R} \, \sigma_{\mu \nu }
q^{\alpha}_{L} )\, , \label{op1}
\end{eqnarray}
where $\alpha$ and $\beta$ are $SU(3)$ colour indices and
${\cal{F}}^{\mu \nu}$ and ${\cal{G}}^{\mu \nu}$ are the field
strength tensors of the electromagnetic and strong interactions,
respectively.
\section{The initial values of the Wilson coefficients.}\label{App2}
The initial values of the Wilson coefficients for the relevant
process in the SM are \cite{Grinstein}
\begin{eqnarray}
C^{SM}_{1,3,\dots 6}(m_W)&=&0 \nonumber \, \, , \\
C^{SM}_2(m_W)&=&1 \nonumber \, \, , \\
C_7^{SM}(m_W)&=&\frac{3 x_t^3-2 x_t^2}{4(x_t-1)^4} \ln x_t+
\frac{-8 x_t^3-5 x_t^2+7 x_t}{24 (x_t-1)^3} \nonumber \, \, , \\
C_8^{SM}(m_W)&=&-\frac{3 x_t^2}{4(x_t-1)^4} \ln x_t+
\frac{-x_t^3+5 x_t^2+2 x_t}{8 (x_t-1)^3}\nonumber \, \, , \\
C_9^{SM}(m_W)&=&-\frac{1}{sin^2\theta_{W}} B(x_t) + \frac{1-4
\sin^2 \theta_W}{\sin^2 \theta_W} C(x_t)-D(x_t)+
\frac{4}{9}, \nonumber \, \, , \\
C_{10}^{SM}(m_W)&=&\frac{1}{\sin^2\theta_W}
(B(x_t)-C(x_t))\nonumber \,\, , \\
C_{Q_i}^{SM}(m_W) & = & 0~~~ i=1,..,10
\end{eqnarray}
and for the additional part due to charged Higgs bosons are
\begin{eqnarray}
C^{H}_{1,\dots 6 }(m_W)&=&0 \nonumber \, , \\
C_7^{H}(m_W)&=& Y^2 \, F_{1}(y_t)\, + \, X Y \,  F_{2}(y_t)
\nonumber  \, \, , \\
C_8^{H}(m_W)&=& Y^2 \,  G_{1}(y_t) \, + \, X Y \, G_{2}(y_t)
\nonumber\, \, , \\
C_9^{H}(m_W)&=&  Y^2 \,  H_{1}(y_t) \nonumber  \, \, , \\
C_{10}^{H}(m_W)&=& Y^2 \,  L_{1}(y_t) \label{CH} \, \, ,
\end{eqnarray}
where
\bea X & = &
\frac{1}{m_{b}}~~~\left(\bar{\xi}^{D}_{N,bb}+\bar{\xi}^{D}_{N,db}
\frac{V_{td}}{V_{tb}} \right) ~~,~~ \nnb \\
Y & = &
\frac{1}{m_{t}}~~~\left(\bar{\xi}^{U}_{N,tt}+\bar{\xi}^{U}_{N,tc}
\frac{V^{*}_{cd}}{V^{*}_{td}} \right) ~~.~~ \eea
Note that the results for model I and II can be obtained from model III
by the following substitutions:
\begin{eqnarray}
Y \rightarrow \cot \beta \, \, & , & \,\, X Y\rightarrow -\cot^2 \beta \,\, for\,\,
 model \, I \nnb \\
Y \rightarrow \cot \beta \, \, & , & \,\, X Y\rightarrow 1 \,\, for\,\, model \, II \, .\nnb
\end{eqnarray}
The NHB effects bring new operators and the corresponding Wilson
coefficients  read as \cite{Iltan2}
\bea
%\!\!\!\!\!\!\!\!\!\!\!\!\!\!\!\!\!\!\!\!\!\!\!\!\!\!\!\!\!\!\!\!\!\!\!\!\!\!\!
%\!\!\!\!\!\!\!\!\!\!\!\!\!\!\!\!\!\!\!\!\!\!\!\!\!\!\!\!\!\!\!\!\!\!\!\!\!\!\!
C^{A^{0}}_{Q_{2}}((\bar{\xi}^{U}_{N,tt})^{3}) & = &
\frac{\bar{\xi}^{D}_{N,\tau \tau}(\bar{\xi}^{U}_{N,tt})^{3} m_{b}
y_t (\Theta_5 (y_t)z_A-\Theta_1 (z_{A},y_t))}{32
\pi^{2}m_{A^{0}}^{2}
m_{t} \Theta_1 (z_{A},y_t) \Theta_5 (y_t)} , \nnb \\
C^{A^{0}}_{Q_{2}}((\bar{\xi}^{U}_{N,tt})^{2})& = &
\frac{\bar{\xi}^{D}_{N,\tau\tau}(\bar{\xi}^{U}_{N,tt})^{2}
\bar{\xi}^{D}_{N,bb}}{32 \pi^{2}  m_{A^{0}}^{2}}\Big{(}
\frac{1}{\Theta_1 (z_{A},y_t) \Theta_1 (z_{A},y_t) \Theta_5
(y_t)}\Big) \nnb
\\ & \cdot & (y_t (\Theta_1 (z_{A},y_t) - \Theta_5 (y_t) (xy+z_A))-
2 \Theta_1 (z_{A},y_t) \Theta_5 (y_t)   \ln [\frac{z_A \Theta_5
(y_t)}{\Theta_1(z_A,y_t)}]) \nnb \eea \bea
C^{A^{0}}_{Q_{2}}(\bar{\xi}^{U}_{N,tt}) &=&
\frac{g^2\bar{\xi}^{D}_{N,\tau\tau}\bar{\xi}^{U}_{N,tt} m_b
x_t}{64 \pi^2 m_{A^{0}}^{2}  m_t } \Bigg{(}\frac{2}{\Theta_5
(x_t)} - \frac{xy x_t+2 z_A}{\Theta_1 (z_{A},x_t)}-2 \ln
[\frac{z_A \Theta_5(x_t)}{ \Theta_1 (z_{A},x_t)}]\nnb \\ &- & x y
x_t y_t(\frac{(x-1) x_t (y_t/z_A-1)-(1+x)y_t)}{(\Theta_6
-(x-y)(x_t -y_t))(\Theta_3 (z_A)+(x-y)(x_t-y_t)z_A)} \nnb \\ & - &
\frac{x (y_t+x_t(1-y_t/z_A))-2 y_t }{\Theta_6 \Theta_3 (z_A)})
\Bigg{)} \nnb \eea \bea
\!\!\!C^{A^{0}}_{Q_{2}}(\bar{\xi}^{D}_{N,bb})& = &
\frac{g^2\bar{\xi}^{D}_{N,\tau\tau}\bar{\xi}^{D}_{N,bb}}{64 \pi^2
m^2_{A^{0}} } \Big{(}1- \frac{x^2_t y_t+2 y (x-1)x_t y_t-z_A
(x^2_t+\Theta_6)}{ \Theta_3 (z_A)} \nnb
\\ & + &
\frac{x^2_t (1-y_t/z_A)}{\Theta_6}+2 \ln [\frac{z_A \Theta_6} {
\Theta_2 (z_{A},x)}] \Big{)}\nnb \eea
\bea
\lefteqn{\!\!\!\!\!\!\!\!\!\!\!\!\!\!\!\!\!\!\!\!\!C^{H^{0}}_{Q_{1}}
((\bar{\xi}^{U}_{N,tt})^{2}) = \frac{g^2 (\bar{\xi}^{U}_{N,tt})^2
m_b m_{\tau} }{64 \pi^2 m^2_{H^{0}} m^2_t } \Bigg{(} \frac{x_t
(1-2 y) y_t}{\Theta_5 (y_t)}+\frac{(-1+2 \cos^2 \theta_W) (-1+x+y)
y_t} {\cos^2 \theta_W \Theta_4 (y_t)} } \nnb
\\ & &
+\frac{z_H (\Theta_1 (z_H,y_t) x y_t + \cos^2 \theta_W \,(-2 x^2
(-1+x_t) y y^2_t+x x_t y y^2_t-\Theta_8 z_H))} {\cos^2 \theta_W
\Theta_1 (z_H,y_t) \Theta_7 }\Bigg{)} ,  \nnb \eea
\bea C^{H^{0}}_{Q_{1}}
 (\bar{\xi}^{U}_{N,tt})
 & = & \frac{g^2 \bar{\xi}^{U}_{N,tt} \bar{\xi}^{D}_{N,bb}
m_{\tau}}{64 \pi^2 m^2_{H^{0}} m_t } \Bigg{(} \frac{(-1+2 \cos^2
\theta_W)\, y_t}{\cos^2 \theta_W \, \Theta_4 (y_t)}- \frac{x_t
y_t}{\Theta_5 (y_t)}+\frac {x_t y_t(x y-z_H)}
{\Theta_1 (z_H,y_t)}  \nnb \\
& + & \frac{(-1+2 \cos^2 \theta_W) y_t z_H}{\cos^2\theta_W
\Theta_7}-2 x_t\, \ln \Bigg{[} \frac{\Theta_5 (y_t) z_H} {\Theta_1
(z_H,y_t)} \Bigg{]} \Bigg{)}  , \label{NHB} \eea
\bea \lefteqn{ C^{H^0}_{Q_{1}}(g^4) =-\frac{g^4 m_b m_{\tau} x_t}
{128 \pi^2 m^2_{H^{0}} m^2_t} \Bigg{(} -1+\frac{(-1+2x)
x_t}{\Theta_5 (x_t) + y (1-x_t)}+
\frac{2 x_t (-1+ (2+x_t) y)}{\Theta_5 (x_t)} } \nnb \\
& & -\frac{4 \cos^2 \theta_W (-1+x+y)+ x_t(x+y)} {\cos^2 \theta_W
\Theta_4 (x_t)} +\frac{x_t (x (x_t (y-2 z_H)-4 z_H)+2 z_H)}
{\Theta_1
(z_H,x_t)} \nnb \\
& & +\frac{y_t ( (-1+x) x_t z_H+\cos^2 \theta_W ( (3 x-y) z_H+x_t
(2 y (x-1)- z_H (2-3 x -y))))}{\cos^2 \theta_W (\Theta_3 (z_H)+x
(x_t-y_t) z_H)} \nnb
\\ & &
+ 2\, ( x_t \ln \Bigg{[} \frac{\Theta_5 (x_t) z_H}{\Theta_1
(z_H,x_t)} \Bigg{]}+ \ln \Bigg{[} \frac{x(y_t-x_t) z_H-\Theta_3
(z_H)} {(\Theta_5 (x_t)+ y (1-x_t) y_t z_H} \Bigg{]} )\Bigg{)}
,\nnb \eea
\bea C^{h_0}_{Q_1}((\bar{\xi}^U_{N,tt})^3) &=&
-\frac{\bar{\xi}^D_{N,\tau\tau} (\bar{\xi}^U_{N,tt})^3 m_b y_t}
{32 \pi^2 m_{h^0}^2 m_t \Theta_1 (z_h,y_t) \Theta_5 (y_t)}
 \Big{(} \Theta_1 (z_h,y_t) (2 y-1) + \Theta_5 (y_t) (2 x-1) z_h \Big{)} \nnb
\eea \bea C^{h_0}_{Q_1}((\bar{\xi}^U_{N,tt})^2) & = &
\frac{\bar{\xi}^D_{N,\tau\tau} (\bar{\xi}^U_{N,tt})^2 } {32 \pi^2
m_{h^0}^2  } \Bigg{(} \frac{ (\Theta_5 (y_t) z_h
(y_t-1)(x+y-1)-\Theta_1 (z_h,y_t)( \Theta_5(y_t)+y_t ) }{\Theta_1
(z_h)\Theta_5(y_t)} \nnb \\ &-& 2 \ln \Bigg{[} \frac{z_h \Theta_5
(y_t)}{\Theta_1 (z_h)} \Bigg{]} \Bigg{)}\nnb \eea
\bea C^{h^0}_{Q_{1}}(\bar{\xi}^{U}_{N,tt}) & = & -\frac{g^2
\bar{\xi}^{D}_{N,\tau\tau}\bar{\xi}^{U}_{N,tt} m_b x_t}{64 \pi^2
m^2_{h^{0}} m_t} \Bigg{(}\frac{2 (-1+(2+x_t) y)}{\Theta_5
(x_t)}-\frac{x_t (x-1)(y_t-z_h)}{\Theta'_2 (z_h)}+2 \ln
\Bigg{[}\frac{z_h \Theta_5 (x_t)}{\Theta_1 (z_h,x_t)} \Bigg{]}
\nnb \\ & + & \frac{x (x_t (y-2 z_h)- 4 z_h)+2 z_h}{\Theta_1
(z_h,x_t)}  -  \frac{(1+x) y_t z_h}{x y x_t y_t+z_h
((x-y)(x_t-y_t)- \Theta_6)} \nnb \\
& + & \frac{\Theta_9 + y_t z_h ( (x-y)(x_t-y_t)-\Theta_6 )(2
x-1)}{z_h \Theta_6 (\Theta_6 -(x-y)(x_t-y_t))}+\frac{x (y_t z_h +
x_t (z_h-y_t))-2 y_t z_h}{\Theta_2 (z_h)} \Bigg{)}, \nnb \eea
\bea C^{h^0}_{Q_{1}}(\bar{\xi}^{D}_{N,bb}) &  = & -\frac{g^2
\bar{\xi}^{D}_{N,\tau\tau}\bar{\xi}^{D}_{N,bb}}{64 \pi^2 m^2_{h^0}
} \Bigg{(}\frac{y x_t y_t (x x^2_t(y_t-z_h)+\Theta_6 z_h
(x-2))}{z_h\Theta_2 (z_h)\Theta_6 }+2 \ln
\Bigg{[}\frac{\Theta_6}{x_t y_t} \Bigg{]} \nnb \\ &+&2 \ln
\Bigg{[}\frac{x_t y_t z_h}{\Theta_2 (z_h)} \Bigg{]} \Bigg{)} \nnb
\eea where \bea \Theta_1 (\omega , \lambda ) & = & -(-1+y-y
\lambda ) \omega -x (y \lambda
+\omega - \omega \lambda ) \nnb \\
\Theta_2 (\omega ) & = &  (x_t +y (1-x_t)) y_t \omega - x x_t (y
y_t+(y_t-1) \omega)   \nnb \\
\Theta^{\prime}_2  (\omega ) & = & \Theta_2 (\omega , x_t \leftrightarrow y_t)    \nnb \\
\Theta_3 (\omega) & = & (x_t (-1+y)-y ) y_t \omega +
x x_t (y y_t+\omega(-1+y_t)) \nnb \\
\Theta_4 (\omega) & = & 1-x +x  \omega  \nnb \\
\Theta_5 (\lambda) & = & x + \lambda (1-x) \nnb \\
\Theta_6  & = & (x_t +y  (1-x_t))y_t +x x_t  (1-y_t) \nnb \\
\Theta_7  & = & (y (y_t -1)-y_t) z_H+x (y y_t + (y_t-1) z_H ) \\
\Theta_8  & = & y_t (2 x^2 (1+x_t) (y_t-1) +x_t (y(1-y_t)+y_t)+x
(2(1-y+y_t) \nnb \\ & + & x_t (1-2 y (1-y_t)-3 y_t))) \nnb \\
\Theta_9  & = & -x^2_t (-1+x+y)(-y_t+x (2 y_t-1)) (y_t-z_h)-x_t
y_t z_h (x(1+2 x)-2 y) \nnb \\ & + & y^2_t (x_t (x^2 -y
(1-x))+(1+x) (x-y) z_h) \nnb \eea and
\begin{eqnarray}
& & x_t=\frac{m_t^2}{m_W^2}~~~,~~~y_t=
\frac{m_t^2}{m_{H^{\pm}}}~~~,~~~z_H=\frac{m_t^2}{m^2_{H^0}}~~~,~~~
z_h=\frac{m_t^2}{m^2_{h^0}}~~~,~~~
z_A=\frac{m_t^2}{m^2_{A^0}}~~~,~~~ \nnb
\end{eqnarray}
The explicit forms of the functions $F_{1(2)}(y_t)$,
$G_{1(2)}(y_t)$, $H_{1}(y_t)$ and $L_{1}(y_t)$ in Eq.(\ref{CH})
are given as
\begin{eqnarray}
F_{1}(y_t)&=& \frac{y_t(7-5 y_t-8 y_t^2)}{72 (y_t-1)^3}+
\frac{y_t^2 (3 y_t-2)}{12(y_t-1)^4} \,\ln y_t \nonumber  \,\, ,
\\
F_{2}(y_t)&=& \frac{y_t(5 y_t-3)}{12 (y_t-1)^2}+ \frac{y_t(-3
y_t+2)}{6(y_t-1)^3}\, \ln y_t \nonumber  \,\, ,
\\
G_{1}(y_t)&=& \frac{y_t(-y_t^2+5 y_t+2)}{24 (y_t-1)^3}+
\frac{-y_t^2} {4(y_t-1)^4} \, \ln y_t \nonumber  \,\, ,
\\
G_{2}(y_t)&=& \frac{y_t(y_t-3)}{4 (y_t-1)^2}+\frac{y_t}
{2(y_t-1)^3} \, \ln y_t  \nonumber\,\, ,
\\
H_{1}(y_t)&=& \frac{1-4 sin^2\theta_W}{sin^2\theta_W}\,\,
\frac{xy_t}{8}\,
\left[ \frac{1}{y_t-1}-\frac{1}{(y_t-1)^2} \ln y_t \right]\nonumber \\
&-& y_t \left[\frac{47 y_t^2-79 y_t+38}{108 (y_t-1)^3}- \frac{3
y_t^3-6 y_t+4}{18(y_t-1)^4} \ln y_t \right] \nonumber  \,\, ,
\\
L_{1}(y_t)&=& \frac{1}{sin^2\theta_W} \,\,\frac{x y_t}{8}\,
\left[-\frac{1}{y_t-1}+ \frac{1}{(y_t-1)^2} \ln y_t \right]
\nonumber  \,\, .
\\
\label{F1G1}
\end{eqnarray}
Finally, the initial values of the coefficients in the model III
are
\begin {eqnarray}
C_i^{2HDM}(m_{W})&=&C_i^{SM}(m_{W})+C_i^{H}(m_{W}) , \nnb \\
C_{Q_{1}}^{2HDM}(m_{W})&=& \int^{1}_{0}dx \int^{1-x}_{0} dy \,
(C^{H^{0}}_{Q_{1}}((\bar{\xi}^{U}_{N,tt})^{2})+
 C^{H^{0}}_{Q_{1}}(\bar{\xi}^{U}_{N,tt})+
 C^{H^{0}}_{Q_{1}}(g^{4})+C^{h^{0}}_{Q_{1}}((\bar{\xi}^{U}_{N,tt})^{3}) \nnb
\\ & + &
 C^{h^{0}}_{Q_{1}}((\bar{\xi}^{U}_{N,tt})^{2})+
 C^{h^{0}}_{Q_{1}}(\bar{\xi}^{U}_{N,tt})+
 C^{h^{0}}_{Q_{1}}(\bar{\xi}^{D}_{N,bb})) , \nnb  \\
 C_{Q_{2}}^{2HDM}(m_{W})&=& \int^{1}_{0}dx \int^{1-x}_{0} dy\,
(C^{A^{0}}_{Q_{2}}((\bar{\xi}^{U}_{N,tt})^{3})+
C^{A^{0}}_{Q_{2}}((\bar{\xi}^{U}_{N,tt})^{2})+
 C^{A^{0}}_{Q_{2}}(\bar{\xi}^{U}_{N,tt})+
 C^{A^{0}}_{Q_{2}}(\bar{\xi}^{D}_{N,bb}))\nnb \\
C_{Q_{3}}^{2HDM}(m_W) & = & \frac{m_b}{m_{\ell} \sin^2 \theta_W}
 (C_{Q_{1}}^{2HDM}(m_W)+C_{Q_{2}}^{2HDM}(m_W)) \nnb \\
C_{Q_{4}}^{2HDM}(m_W) & = & \frac{m_b}{m_{\ell} \sin^2 \theta_W}
 (C_{Q_{1}}^{2HDM}(m_W)-C_{Q_{2}}^{2HDM}(m_W)) \nnb \\
C_{Q_{i}}^{2HDM} (m_W) & = & 0\,\, , \,\, i=5,..., 10. \label{CiW}
\end{eqnarray}
Here, we present $C_{Q_{1}}$ and $C_{Q_{2}}$ in terms of the
Feynman parameters $x$ and $y$ since the integrated results are
extremely large. Using these initial values, we can calculate the
coefficients $C_{i}^{2HDM}(\mu)$ and $C^{2HDM}_{Q_i}(\mu)$ at any
lower scale in the effective theory with five quarks, namely
$u,c,d,s,b$ similar to the SM case \cite{Misiak}-\cite{Buras}.

The Wilson  coefficients playing  the essential role in this
process are $C_{7}^{2HDM}(\mu)$, $C_{9}^{2HDM}(\mu)$,
$C_{10}^{2HDM}(\mu)$, $C^{2HDM}_{Q_1}(\mu )$ and
$C^{2HDM}_{Q_2}(\mu )$. For completeness, in the following we give
their explicit expressions.
\begin{eqnarray}
C_{7}^{eff}(\mu)&=&C_{7}^{2HDM}(\mu)+ Q_d \, (C_{5}^{2HDM}(\mu) +
N_c \, C_{6}^{2HDM}(\mu))\nonumber \, \, , \label{C7eff}
\end{eqnarray}
where the LO  QCD corrected Wilson coefficient $C_{7}^{LO,
2HDM}(\mu)$ is given by
\begin{eqnarray}
C_{7}^{LO, 2HDM}(\mu)&=& \eta^{16/23} C_{7}^{2HDM}(m_{W})+(8/3)
(\eta^{14/23}-\eta^{16/23}) C_{8}^{2HDM}(m_{W})\nonumber \,\, \\
&+& C_{2}^{2HDM}(m_{W}) \sum_{i=1}^{8} h_{i} \eta^{a_{i}} \,\, ,
\label{LOwils}
\end{eqnarray}
and $\eta =\alpha_{s}(m_{W})/\alpha_{s}(\mu)$, $h_{i}$ and $a_{i}$
are the numbers which appear during the evaluation \cite{Buras}.

$C_9^{eff}(\mu)$ contains a perturbative part and a part coming
from LD effects due to conversion of the real $\bar{c}c$ into
lepton pair $\ell^+ \ell^-$:
\begin{eqnarray}
C_9^{eff}(\mu)=C_9^{pert}(\mu)+ Y_{reson}(s)\,\, ,
\label{C9efftot}
\end{eqnarray}
where
\begin{eqnarray}
C_9^{pert}(\mu)&=& C_9^{2HDM}(\mu) \nonumber
\\ &+& h(z,  s) [ 3 C_1(\mu) + C_2(\mu) + 3 C_3(\mu) +
C_4(\mu) + 3 C_5(\mu) + C_6(\mu) \nonumber \\&+&\lambda_u(3C_1 +
C_2) ] -  \frac{1}{2} h(1, s) \left( 4 C_3(\mu) + 4 C_4(\mu)
+ 3 C_5(\mu) + C_6(\mu) \right)\nnb \\
&- &  \frac{1}{2} h(0,  s) \left[ C_3(\mu) + 3 C_4(\mu) -\lambda_u
(6 C_1(\mu)+2C_2(\mu)) \right] \\&+& \frac{2}{9} \left( 3 C_3(\mu)
+ C_4(\mu) + 3 C_5(\mu) + C_6(\mu) \right) \nonumber \,\, ,
\end{eqnarray}
and
\begin{eqnarray}
Y_{reson}(s)&=&-\frac{3}{\alpha^2_{em}}\kappa \sum_{V_i=\psi_i}
\frac{\pi \Gamma(V_i\rightarrow \ell^+
\ell^-)m_{V_i}}{q^2-m_{V_i}+i m_{V_i}
\Gamma_{V_i}} \nonumber \\
&\times & [ (3 C_1(\mu) + C_2(\mu) + 3 C_3(\mu) + C_4(\mu) + 3
C_5(\mu) + C_6(\mu))\nnb\\ &+&\lambda_u(3C_1(\mu)+C_2(\mu))]\, .
 \label{Yresx}
\end{eqnarray}
In Eq.(\ref{C9efftot}), the functions $h(u, s)$ are given by
\begin{eqnarray}
h(u, s) &=& -\frac{8}{9}\ln\frac{m_b}{\mu} - \frac{8}{9}\ln u +
\frac{8}{27} + \frac{4}{9} x \\
& & - \frac{2}{9} (2+x) |1-x|^{1/2} \left\{\begin{array}{ll}
\left( \ln\left| \frac{\sqrt{1-x} + 1}{\sqrt{1-x} - 1}\right| -
i\pi \right), &\mbox{for } x \equiv \frac{4u^2}{ s} < 1 \nonumber \\
2 \arctan \frac{1}{\sqrt{x-1}}, & \mbox{for } x \equiv \frac
{4u^2}{ s} > 1,
\end{array}
\right. \\
h(0,s) &=& \frac{8}{27} -\frac{8}{9} \ln\frac{m_b}{\mu} -
\frac{4}{9} \ln s + \frac{4}{9} i\pi \,\, , \label{hfunc}
\end{eqnarray}
with $u=\frac{m_c}{m_b}$. The phenomenological parameter $\kappa$
in Eq. (\ref{Yresx}) is taken as $2.3$. In Eqs. (B.10) and
(\ref{Yresx}), the contributions of the coefficients $C_1(\mu)$,
...., $C_6(\mu)$ are due to the operator mixing.

Finally, the Wilson coefficients $C_{Q_1}(\mu)$ and $C_{Q_2}(\mu
)$ are given by \cite{Dai} \beq C_{Q_i}(\mu
)=\eta^{-12/23}\,C_{Q_i}(m_W)~,~i=1,2~. \eeq
%
%\end{appendix}
%
%\listoftables \listoffigures

\end{document}